%% file: main.tex
\documentclass[11pt, reqno]{amsart}

\usepackage[letterpaper,hmargin=1in,vmargin=1in]{geometry}

\parskip 	\smallskipamount

\usepackage{fullpage,url,amssymb,enumerate,colonequals,mathrsfs,MnSymbol,extarrows,lscape,empheq,color,float}
\usepackage[all,cmtip]{xy}
\usepackage[OT2,T1]{fontenc}

\restylefloat{table}

\usepackage{subcaption}
\usepackage{listings}
\usepackage{tikz}
\usepackage{xstring}
\usetikzlibrary{calc}
\newcommand{\Size}{1cm}

\tikzset{Square/.style={
    inner sep=0pt,
    text width=\Size, 
    minimum size=\Size,
    draw=black,
    fill=yellow!20,
    align=center
    }
}
\usetikzlibrary{decorations.markings}
\tikzstyle{vertex}=[circle, draw, inner sep=0pt, minimum size=6pt]
\newcommand{\vertex}{\node[vertex]}

\usepackage{amsmath}
\DeclareMathOperator*{\argmax}{arg\,max}

\usepackage{verbatim}

\theoremstyle{definition}

\newtheorem{theorem}{Theorem}
\newtheorem{lemma}{Lemma}
\newtheorem{proposition}{Proposition}
\newtheorem{corollary}{Corollary}
\newtheorem{definition}{Definition}
\newtheorem{remark}{Remark}

\usepackage[
    colorlinks=true,
    citecolor=blue,
    linkcolor=blue,
    linktocpage
]{hyperref}

\begin{document}

\title{Upper Bounds for All and Max-gain Policy Iteration Algorithms on Deterministic MDPs}

\author{Ritesh Goenka, Eashan Gupta$^\dagger$, Sushil Khyalia$^\dagger$, Pratyush Agarwal$^*$, Mulinti Shaik Wajid$^*$, and Shivaram Kalyanakrishnan}

\date{\today}



\address{
    Ritesh Goenka \\
    Mathematical Institute \\
    University of Oxford \\
    Oxford \\
    UK.
}

\email{goenka@maths.ox.ac.uk}

\address{
    Eashan Gupta \\
    Department of Computer Science \\
    University of Illinois at Urbana-Champaign \\
    Champaign \\
    Illinois \\
    USA.
}

\email{eashang2@illinois.edu}

\address{
    Sushil Khyalia \\
    Machine Learning Department \\
    Carnegie Mellon University \\
    Pittsburgh \\
    Pennsylvania \\
    USA.
}

\email{skhyalia@andrew.cmu.edu}

\address{
    Pratyush Agarwal \\
    Department of Computer Science \\
    Stanford University \\
    Stanford \\
    California \\
    USA.
}

\email{prat1019@stanford.edu}

\address{
    Mulinti Shaik Wajid \\
    Department of Computer Science and Engineering \\
    Indian Institute of Technology Bombay \\
    Mumbai \\
    Maharashtra \\
    India.
}

\email{wajid@cse.iitb.ac.in}

\address{
    Shivaram Kalyanakrishnan \\
    Department of Computer Science and Engineering \\
    Indian Institute of Technology Bombay \\
    Mumbai \\
    Maharashtra \\
    India.
}

\email{shivaram@cse.iitb.ac.in}

\thanks{$^\dagger$ and $^*$ indicate equal contribution.}

\begin{abstract}
    Policy Iteration (PI) is a widely used family of algorithms to compute optimal policies for Markov Decision Problems (MDPs). We derive upper bounds on the running time of PI on Deterministic MDPs (DMDPs): the class of MDPs in which every state-action pair has a unique next state. Our results include a non-trivial upper bound that applies to the entire family of PI algorithms; another to all ``max-gain'' switching variants; and affirmation that a conjecture regarding Howard's PI on MDPs is true for DMDPs. Our analysis is based on certain graph-theoretic results, which may be of independent interest.
\end{abstract}

\maketitle

\tableofcontents

\input{introduction}
\input{background}
\input{cycles}
\input{dmdp}
\input{twostate}
\input{conclusion}

\bibliography{references}
\bibliographystyle{amsplain}

\end{document}

%% file: introduction.tex
\section{Introduction}
\label{sec:introduction}

A Markov Decision Problem (MDP) (Puterman~\cite{Puterman}) is an abstraction of a decision-making task in which the effect of any given \textit{action} from any given \textit{state} is a stochastic transition to a next state, coupled with a numeric reward. A \textit{policy} (taken in this article to be Markovian and deterministic) for an MDP specifies the action to take from each state. The utility of a policy is usually taken as some form of the expected \textit{long-term} reward it yields. A common definition of long-term reward (also the one we use in this article) is as a discounted sum of the individual rewards over an infinite horizon. For an MDP with a finite number of states and actions, the set of policies is also finite, and this set necessarily contains an \textit{optimal} policy, which maximises the expected long-term reward starting from each state in the MDP (Bellman~\cite{Bellman}). For a given MDP---specified by its sets of states and actions, transition probabilities, rewards, and discount factor---the desired solution is an optimal policy for the MDP.

Policy Iteration (PI) (Howard~\cite{Howard}) is a widely-used family of algorithms to solve MDPs. A PI algorithm is initialised with some arbitrary policy, and iterates through a sequence of policies that is guaranteed to terminate in an optimal policy. In each iteration, a set of ``improving'' actions is identified for each state. Any policy obtained by switching one or more of the current policy's actions to improving actions is guaranteed to dominate the current policy, and hence can be selected as the subsequent iterate. A policy with no improving actions is guaranteed to be optimal. Algorithms from the PI family are distinguished by their choice of improving actions for switching to at each step. An alternative perspective of PI algorithms emerges by considering a linear program $P_{M}$ induced by the input MDP $M$ (Puterman~\cite[see Section 6.9.1]{Puterman}). The vertices of the feasible polytope of $P_{M}$ are in bijective correspondence with policies for $M$. PI algorithms restricted to changing the action only at a single state essentially perform a Simplex update on the feasible polytope. On the other hand, the generic PI update, which could involve changing actions at multiple states, amounts to a block-pivoting step.

A Deterministic MDP (DMDP) is an MDP in which the transitions are all deterministic. In other words, for every state $s$ and action $a$ in a DMDP, there is a unique state $s^{\prime}$ which is reached whenever $a$ is taken from $s$. With the challenge of stochasticity removed, DMDPs would appear to be an easier class of problems to solve than MDPs---and several results affirm this intuition. For example, solving MDPs has been established to be $P$-complete, whereas solving DMDPs is in $NC$ (Papadimitriou~\cite{PapadimitriouTsitsiklis}). Moreover, it has been established that DMDPs can be solved in \textit{strongly polynomial} time: this means that if any arithmetic or relational operation can be performed in constant time, regardless of the size of the operands, then the total number of such operations required is at most a polynomial in the number of states and actions, with no dependence on other input parameters. Madani~\cite{Madani} proposes a specialised algorithm for DMDPs that enjoys a strongly polynomial upper bound, while Post and Ye~\cite{PostYe} establish that the classical ``max-gain'' variant of the Simplex algorithm (an instance of PI) runs in strongly polynomial time on DMDPs.

By upper-bounding the complexity of specific algorithms on DMDPs, the preceding results indirectly upper-bound the running time of the \textit{best} algorithm from some corresponding class of algorithms. In this article, we adopt a complementary perspective. We derive running-time upper bounds that apply to the entire family of PI algorithms, and hence to the \textit{worst} among them, when run on DMDPs. We also provide similar analysis when action selection follows the ``max-gain'' rule.

\subsection{Results.} Consider a finite DMDP with $n \geq 2$ states and $k \geq 2$ actions, with the convention that each of the $k$ actions is available from each state. The total number of policies, $k^{n}$, is a trivial upper bound on the number of iterations taken by any PI algorithm. We show that the number of iterations taken by \textit{every} PI algorithm on every $n$-state, $k$-action DMDP is at most 
\begin{equation}
\label{eqn:all-ub}
    \text{poly}(n, k) \cdot \alpha(k)^{n}, \text{ where } \alpha(k) = \frac{k-1 + \sqrt{(k-1)^2 + 4}}{2}.
\end{equation}
In turn, this quantity is upper-bounded by $\text{poly}(n, k) \cdot \left(k - \frac{3 - \sqrt{5}}{2}\right)^{n} = \text{poly}(n, k) \cdot (k - 0.3819\dots)^{n}$.

The significance of \eqref{eqn:all-ub} is most apparent for the special case of $k = 2$ actions. Melekopoglou and Condon~\cite{MelekopoglouCondon} have constructed an $n$-state, $2$-action MDP on which a particular variant of PI visits all the $2^{n}$ policies. On the other hand, our result establishes that no PI algorithm can exceed $\text{poly}(n) \cdot \left(\frac{1 + \sqrt{5}}{2}\right)^{n}$ iterations on any $n$-state, $2$-action DMDP. Another interesting consequence of \eqref{eqn:all-ub} concerns Howard's PI~\cite{Howard}, arguably the most commonly used variant of the PI family. In Howard's PI, a switch is necessarily made at every state with improving actions. On $n$-state, $2$-action MDPs, the tightest upper bound known for this algorithm is $O(\frac{2^{n}}{n})$ iterations (Mansour and Singh~\cite{MansourSingh}, Hollanders~\cite{Hollanders}), which is only a linear improvement over the trivial upper bound. It has been conjectured that the number of iterations of Howard's PI on $n$-state, $2$-action MDPs grows at most at the rate of the Fibonacci numbers (Hansen ~\cite{Hansen}, Gerencs\'{e}r \textit{et al.}~\cite{Gerencser}). The upper bound in \eqref{eqn:all-ub} validates this conjecture, albeit up to a polynomial factor, for the special case of DMDPs.

When a state has multiple improving actions, one common rule to select the action to switch to is based on the actions' ``gains''. The gain of an action $a$ is the difference in utility arising from replacing the current policy with $a$ for only the very first time step. Switching to an action with the maximum gain (``max-gain'')
is a key ingredient of the strongly polynomial upper bound for the Simplex method on DMDPs (Post and Ye~\cite{PostYe}). Max-gain action selection has also been observed to be efficient in practice with other PI variants (Taraviya and Kalyanakrishnan~\cite{TaraviyaKalyanakrishnan}). We obtain a smaller upper bound than \eqref{eqn:all-ub} for sufficiently large $k$ when we restrict the PI algorithm to perform max-gain action selection (while the algorithm is still free to select on which \textit{states} to switch actions). In particular, we show that the total number of iterations for max-gain variants of PI is at most 
\begin{equation}
\label{eqn:maxgain-ub}
    \text{poly}(n, k) \cdot \left(\left(1 + O\left(\frac{\log k}{k}\right)\right)\frac{k}{e}\right)^{n}.
\end{equation}

The upper bounds in \eqref{eqn:all-ub} and \eqref{eqn:maxgain-ub} follow from the solution of two graph-theoretic problems, which constitute the technical core of this article. Every DMDP $M$ induces a directed multigraph $G_{M}$ in which the vertices are the states, and the edges are the actions in $M$. In turn, each policy induces a subgraph of $G_{M}$, with the restriction of having a single outgoing edge from each vertex. For each state (equivalently vertex) $s$, the long-term reward accrued by a policy is fully determined by a directed path starting at $s$, and a directed cycle that the path reaches. Since each PI iterate dominates the preceding one, it follows that no ``path-cycle'' can reappear after once having been replaced. We obtain novel bounds on two structural properties of $G_{M}$, which are defined in terms of path-cycles in $G_{M}$. Correspondingly we obtain upper bounds on the running time of PI algorithms: \eqref{eqn:all-ub} for arbitrary PI variants, and \eqref{eqn:maxgain-ub} for variants that perform ``max gain'' action selection.

As mentioned earlier, the  lower bound of $2^{n}$ iterations shown by Melekopglou and Condon~\cite{MelekopoglouCondon} for $n$-state, $2$-action MDPs implies a clear separation with $n$-state, $2$-action DMDPs, for which \eqref{eqn:all-ub} guarantees that no PI algorithm takes more than $\text{poly}(n) \cdot \left(\frac{1 + \sqrt{5}}{2}\right)^{n}$ iterations. As yet, it is not apparent if such a separation between DMDPs and MDPs also holds for $k \geq 3$ actions. More to the point, there are no known lower bounds for PI on MDPs that exceed the upper bounds in \eqref{eqn:all-ub} and \eqref{eqn:maxgain-ub} for DMDPs when $k \geq 3$. As a step towards scoping out the gap between DMDPs and MDPs, we investigate the special case of $n = 2$: that is, of MDPs with $2$ states and $k \geq 2$ actions. In this case, we do obtain a separation between DMDPs and MDPs, both for arbitrary and for max-gain action selection. We construct a $2$-state, $k$-action MDP on which a PI variant visits all $k^{2}$ policies, and another $2$-state, $k$-action MDP on which a max-gain PI variant evaluates $k$ policies. We also show on the other hand that on every $2$-state, $k$-action MDP, every PI variant visits at most $\frac{k^{2}}{2} + 2k - 1$ policies, and every max-gain PI variant visits at most $7$ policies. These results motivate the pursuit of tighter upper bounds for PI on DMDPs, along with tighter lower bounds on MDPs, for the general case of $n \geq 3$.

The article is organised as follows. We provide requisite definitions and background along with some related work in Section~\ref{sec:background}, before presenting our results in Sections~\ref{sec:cycles} and \ref{sec:DMDPs}. The reader prepared to accept our graph-theoretic results without proof may skip Section~\ref{sec:cycles} and proceed directly to Section~\ref{sec:DMDPs}, in which we use these results to prove the upper bounds \eqref{eqn:all-ub} and \eqref{eqn:maxgain-ub}. Our specialised analysis of $2$-state MDPs follows in Section~\ref{sec:2statebounds}.

%% file: background.tex
\section{Policy iteration}
\label{sec:background}

We begin by defining MDPs, and thereafter describe the PI family of algorithms for solving them. Next, we present known upper bounds on the number of steps taken by PI to find an optimal policy.\\

\subsection{Markov Decision Problems.}

\begin{definition}
An MDP is a $5$-tuple $(S, A, T, R, \gamma)$, where $S$ is a set of states; $A$ is a set of actions; $T: S \times A \times S \to [0, 1]$ is the transition function with $T(s, a, s')$ being the probability of reaching state $s^{\prime} \in S$ by taking action $a \in A$ from state $s \in S$ (hence $\sum_{s^{\prime}} T(s, a, s^{\prime}) = 1$); $R: S \times A \to \mathbb{R}$ is the reward function with $R(s, a)$ being the expected reward obtained on taking action $a \in A$ from state $s \in S$; and $\gamma \in [0, 1)$ is the discount factor.
\end{definition}

Given an MDP $M = (S, A, T, R, \gamma)$, a \textit{policy} (assumed deterministic and Markovian) $\pi: S \to A$ specifies an action $a \in A$ for each state $s \in S$. We denote the set of all policies for a given MDP $M$ by $\Pi_M$ (or just $\Pi$ if the underlying MDP $M$ is evident from the context). In this work, we assume that $S$ and $A$ are finite, with $|S| = n \geq 2$ and $|A| = k \geq 2$. Consequently, $\Pi$ is also finite, and contains $k^{n}$ policies.

\begin{definition}
    For policy $\pi \in \Pi$, the \textit{value function} $V^\pi: S \to \mathbb{R}$ gives the expected infinite discounted reward obtained by starting from each state $s \in S$ and following the policy $\pi$. Let $s_{0}, a_{0}, r_{0}, s_{1}, a_{1}, r_{1}, \dots$ be the state-action-reward trajectory generated by $M$ over time (the subscript indicates the time step). Then for $s \in S$,
    \begin{equation*}
        V^\pi(s) := \mathbb{E}\left[ \sum_{t=0}^\infty \gamma^t R(s_t, a_t)\right],
    \end{equation*}
    where $s_0 = s$, and for $t \geq 0$, $a_t = \pi(s_t), s_{t + 1} \sim T(s_{t}, a_{t})$. The \textit{action value function} $Q^\pi: S \times A \to \mathbb{R}$ applied to $s \in S, a \in A$ is the expected infinite discounted reward obtained by starting from $s$ taking $a$, and thereafter following $\pi$. Finally, the \textit{gain function} $\rho^\pi: S \times A \to \mathbb{R}$ provides the difference between $Q^{\pi}$ and $V^{\pi}$.
\end{definition}

All three functions, $V^\pi$, $Q^\pi$, and $\rho^\pi$, can be computed efficiently (in $\text{poly}(n , k)$ arithmetic operations) by solving the following equations for $s \in S, a \in A$. 
\begin{align*}
V^\pi(s) &= R(s, \pi(s)) + \gamma \sum_{s' \in S} T(s,\pi(s),s') V^\pi(s').\\
Q^\pi(s, a) &= R(s, a) + \gamma \sum_{s' \in S} T(s,a,s') V^\pi(s').\\
\rho^\pi(s,a) &= Q^\pi(s,a) - V^\pi(s).
\end{align*}
The first set of equations above, used to compute $V^{\pi}$ for some fixed policy $\pi \in \Pi$, are called the Bellman equations for $\pi$.

We now define relations $\preceq$ and $\prec$ to compare policies in $\Pi$.

\begin{definition}
For $\pi_1, \pi_2 \in \Pi$, $\pi_1 \preceq \pi_2$ if $V^{\pi_1}(s) \le V^{\pi_2}(s)$ for all $s \in S$. Moreover, $\pi_{1} \prec \pi_{2}$ if $\pi_1 \preceq \pi_2$ and additionally $V^{\pi_1}(s) < V^{\pi_2}(s)$ for some state $s \in S$.
\end{definition}

\begin{definition}
An policy $\pi^{\star} \in \Pi$ is called an \textit{optimal} policy if $\pi \preceq \pi^{\star}$ for all $\pi \in \Pi$.
\end{definition}

\subsection{Policy improvement.}

\begin{definition}
For policy $\pi \in \Pi$, the \textit{improvable set} $I^\pi$ is defined as the set of state-action pairs $(s, a) \in S \times A$ such that $\rho^{\pi}(s, a) > 0$. A set $I \subseteq I^\pi$ is said to be a \textit{valid} improvement set for $\pi$ if for each $s \in S$, there exists at most one action $a \in A$ such that $(s, a) \in I$, and moreover, $|I| \geq 1$.
\end{definition}

\begin{definition}
Suppose that for policy $\pi \in \Pi$, the set $I^{\pi}$ is non-empty. Fix an arbitrary, valid improvement set $I \subseteq I^\pi$. Consider policy $\pi' \in \Pi$, given by
\begin{equation}
\label{eqn:policyimprovement}
    \pi'(s) = \begin{cases} 
                a, & \text{if }(s,a) \in I, \\
                \pi(s), & \text{otherwise}.
            \end{cases}
\end{equation}
Then $\pi^{\prime}$ is called a \textit{locally-improving} policy of $\pi$. The operation of obtaining $\pi^{\prime}$ from $\pi$, by switching to corresponding actions in the improvement set $I$, is called \textit{policy improvement}.
\end{definition}

Notice that if $|I^{\pi}| > 1$, there are multiple possible choices of valid improvement sets $I \subseteq I^{\pi}$. The well-known policy improvement theorem, stated below, provides a guarantee that applies to every such choice of $I$.

\begin{theorem}
\label{thm:policy-improvement}
Fix $\pi \in \Pi$. (1) If $I^{\pi} = \emptyset$, then
$\pi$ is an optimal policy. (2) If $I^{\pi} \neq \emptyset$, let $\pi^{\prime} \in \Pi$ be obtained from policy improvement to $\pi$ using any valid improvement set $I \subseteq I^{\pi}$. Then $\pi \prec \pi^{\prime}$, and moreover, for $s \in S$ such that $\pi^{\prime}(s) \neq \pi(s)$, we have $V^{\pi^{\prime}}(s) > V^{\pi}(s)$ .
\end{theorem}

We omit the proof of the theorem, which is verifiable from standard references (Puterman~\cite[see Section 6.4.2]{Puterman}, Szepesv{\'a}ri~\cite[see Appendix A.2]{Szepesvari}).
The theorem establishes the existence of an optimal policy for every MDP. Indeed ``solving'' an MDP amounts to computing an optimal policy for it. Although there are many possible solution techniques, a natural approach is evident from the theorem itself: to iterate through policies. Algorithms from the PI family proceed through a sequence of policies $\pi_{0} \to \pi_{1} \to \dots \to \pi_{\ell}$, wherein $\pi_{0} \in \Pi$ is an arbitrary initial policy; $\pi_{\ell}$ for some $\ell \geq 1$ is an optimal policy; and for $0 \leq i < \ell$, $\pi_{i + 1}$ is obtained from policy improvement on $\pi_{i}$.

\begin{figure}[b]
\centering

\begin{subfigure}{.3\textwidth}

\[\begin{tikzpicture}[x=1.75cm, y=1.75cm,
    every edge/.style={
        draw,
        postaction={decorate,
                    decoration={markings,mark=at position 0.5 with {\arrow[scale=2]{>}}}
                   }
        },
    every loop/.style={},
    el/.style = {inner sep=2pt, align=left, sloped},
]

\vertex[minimum width=0.6cm,circle,inner sep=0pt] (v1) at (0,1) {$s_0$};
\vertex[minimum width=0.6cm,circle,inner sep=0pt] (v2) at  (-0.866,-0.5) {$s_1$};
\vertex[minimum width=0.6cm,circle,inner sep=0pt] (v3) at  (0.866,-0.5) {$s_2$};

\path
    (v1) edge[in=75,out=15,loop,dashed,color=red] node[el,above=1mm] {$1,3$} (v1)
    (v1) edge[in=105,out=165,loop] node[el,above=1mm] {$0.25,2$} (v1)
    (v1) edge[bend right=10] node[el,above=1mm] {$0.5,3$} (v2)
    (v1) edge[bend left=10] node[el,above=1mm] {$0.25,3$} (v3)
    (v2) edge[bend left=10,dashed,color=red] node[el,above=1mm] {$0.5,2$} (v3)
    (v2) edge[bend right=10,dashed,color=red] node[el,below=1mm] {$0.5,2$} (v1)
    (v3) edge[bend left=10,dashed,color=red] node[el,below=1mm] {$0.5,3$} (v1)
    (v3) edge[bend left=10,dashed,color=red] node[el,below=1mm] {$0.5,1$} (v2)
    (v3) edge[bend left=50] node[el,below=1mm] {$1,3$} (v2)
    (v2) edge[in=-120,out=-60,loop] node[el,below=1mm] {$1,3$} (v2)

;
\end{tikzpicture}\]
\end{subfigure}
\begin{subfigure}{.3\textwidth}
\[\begin{tikzpicture}[x=2cm, y=2cm,
    every edge/.style={
        draw,
        postaction={decorate,
                    decoration={markings,mark=at position 0.47 with {\arrow[scale=2]{>}}}
                   }
        }
]

\vertex (v000) at  (0,-1.061) [label=below:$000$]{};
\vertex (v001) at  (-0.707,-0.354) [label={[label distance=1mm]270:$001$}]{};
\vertex (v010) at  (0.707,-0.354) [label={[label distance=1mm]270:$010$}]{};
\vertex (v011) at  (0,0.354) [label={[label distance=0mm]90:$011$}]{};
\vertex (v100) at  (0,-0.354) [label={[label distance=0mm]270:$100$}]{};
\vertex (v101) at  (-0.707,0.354) [label={[label distance=1mm]90:$101$}]{};
\vertex (v110) at  (0.707,0.354) [label={[label distance=1mm]90:$110$}]{};
\vertex (v111) at  (0,1.061) [label=above:$111$]{};

\path

(v000) edge (v010)
(v000) edge (v001)

(v001) edge (v011)

(v010) edge (v011)

(v100) edge (v000)
(v100) edge (v101)
(v100) edge (v110)

(v101) edge (v001)
(v101) edge (v111)

(v110) edge (v010)
(v110) edge (v111)

(v111) edge (v011)

(v101) edge[dashed] (v011)
(v110) edge[dashed] (v011)
(v100) edge[dashed] (v001)
(v100) edge[dashed] (v010)
(v100) edge[dashed] (v011)

(v000) edge[bend right=30,dashed] (v011)
(v100) edge[bend left=30,dashed] (v111)
;
\end{tikzpicture}\]
\end{subfigure}
\caption{An example of a $3$-state $2$-action MDP with $\gamma=0.9$ (left) and the corresponding policy improvement digraph (right). In the left figure, the dashed (red) edges and solid (black) edges correspond to actions $0$ and $1$, respectively. Each transition in the MDP is marked with its (transition probability, reward) pair. In the right figure, the solid and dashed edges correspond to policy improvement steps that switch action on a single state and multiple states, respectively.}
\label{fig:mdp-auso}
\end{figure}
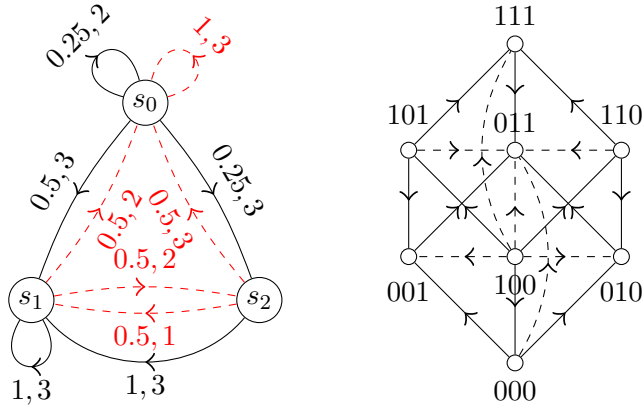

Given any MDP $M$, we can now define its Policy Improvement Directed Acyclic Graph (PI-DAG) in the following manner: it is a directed graph with $\Pi$ as its set of vertices, and there is an edge from $\pi$ to $\pi'$ if $\pi'$ can be obtained from $\pi$ by a single step of policy improvement. Note that the resulting digraph is acyclic since policy improvement always yields a strictly dominating policy. Figure~\ref{fig:mdp-auso} shows an example MDP and its PI-DAG. Note that directed paths starting at some vertex and ending at a sink vertex in the PI-DAG are in bijective correspondence with the possible trajectories of PI algorithms on the MDP $M$.

\subsection{Switching rules.} Variants from the PI family are distinguished by their ``switching rule''---in other words their choice of valid improvement set $I \subseteq I^{\pi}$ to improve the current policy $\pi$. In principle, this choice can depend on the entire sequence of policies visited yet, along with any accompanying information gathered from each iteration. However, most common variants of PI are ``memoryless'': that is, they select the improvement set $I \in I^{\pi}$ solely based on $I^{\pi}$, and sometimes with additional knowledge of $\rho^{\pi}(s, a)$ for $(s, a) \in I^{\pi}$. For our purposes, it is convenient to view switching rules as a sequence of two steps: the first to select which \textit{states} will be given new actions, and the second to select improving actions for each of these states. Concretely, let $S^{+}(\pi)$ denote the set of all states $s \in S$ for which there exists some action $a \in A$ such that $(s, a) \in I^{\pi}$. For each state $s \in S^{+}(\pi)$, let $A^{+}(\pi, s)$ be the set of actions $a \in A$ such that $(s, a) \in I^{\pi}$. Any switching rule must select a non-empty subset $S_{\text{switch}} \subseteq S^{+}(\pi)$, and for each $s \in S_{\text{switch}}$, select an action $a_{\text{switch}}(s) \in A^{+}(\pi, s)$. Note that in general, $S_{\text{switch}}$ and $a_{\text{switch}}(\cdot)$ can both be random. Also, note that $a_{\text{switch}}(s)$ is trivially determined for each $s \in S_{\text{switch}}$ when the MDP has only $k = 2$ actions.

In our upcoming analysis of the number of iterations taken by PI on DMDPs, we place no restriction on how $S_{\text{switch}}$ is selected from $S^{+}(\pi)$. However, we consider two distinct settings for action selection.
(1) With \textit{arbitrary} action selection, there is no restriction on how $a_{\text{switch}}(s)$ is selected from $A^{+}(\pi, s)$ for $s \in S_{\text{switch}}$. (2) Under \textit{max-gain} action selection, we have 
$a_{\text{switch}}(s) \in \argmax_{a \in A^{+}(\pi, s)} \rho^{\pi}(s, a)$ for $s \in S_{\text{switch}}$, with arbitrary tie-breaking. The max-gain action selection rule is used widely in practice. To the best of our knowledge, existing upper bounds on the complexity of PI on MDPs (presented shortly) all assume some constraint on the state selection step in the switching rule. Since we place no such restriction, our upper bounds when action selection is arbitrary apply to \textit{every} PI algorithm, including those whose switching choices depend on memory and additional information. 

\subsection{Known results on complexity.}
We briefly review results on the running time of PI, restricting ourselves to bounds that depend only on the number of states $n \geq 2$ and actions $k \geq 2$ in the input MDP. Arguably the most common variant from the PI family is Howard's PI \cite{Howard}, under which $S_{\text{switch}} = S^{+}(\pi)$. Mansour and Singh~\cite{MansourSingh} show an upper bound of $O(k^{n}/n)$ iterations when Howard's PI is coupled with arbitrary action selection; Taraviya and Kalyanakrishnan \cite{TaraviyaKalyanakrishnan} obtain a tighter bound of $\left(O(\sqrt{k \log k})\right)^n$ iterations when action selection is random. Mansour and Singh \cite{MansourSingh} also propose a randomised PI variant in which $S_{\text{switch}}$ is chosen uniformly at random from among the non-empty subsets of $S^{+}(\pi)$. They give an upper bound of $O\left(\left((1+\frac{2}{\log k})\frac{k}{2}\right)^n\right)$ iterations (with high probability) when action selection is arbitrary. Their bound of $O(2^{0.78 n})$ expected iterations for the special case of $k = 2$ was subsequently improved by Hansen \textit{et al.}~\cite{HansenPatZwick} to $O(n^5)\left(\frac{3}{2}\right)^n$.

PI variants that switch only a single state in each iteration (that is, which enforce $|S_{\text{switch}}| = 1$) may be interpreted as variants of the Simplex algorithm, being run on an LP $P_{M}$ induced by the input MDP $M$. Vertices in the feasible polytope of $P_{M}$ are in bijective correspondence with the set of policies $\Pi$; single switch policy improvements amount to shifting to a neighbouring vertex that increases the objective function, which at the vertex corresponding to policy $\pi \in \Pi$ is $\sum_{s \in S} V^{\pi}(s)$. Suppose the set of states $S$ is indexed; without loss of generality take $S = \{1, 2, \dots, n\}$. Kalyanakrishnan et al. \cite{KalMisGop} show an upper bound of $(2 + \ln (k-1))^n$ expected iterations for a variant of PI in which $S_{\text{switch}} = \{\max_{s \in S^{+}(\pi)} s\}$, and action selection is random. Interestingly, Melekopoglou and Condon \cite{MelekopoglouCondon} show that the same rule results in a policy improvement sequence of length $2^{n}$ for an $n$-state, $2$-action MDP. Among deterministic variants of PI, the best known upper bound is $\text{poly}(n,k) \cdot k^{0.7207 n}$ iterations (Gupta and Kalyanakrishnan~\cite{GupKal}), for a variant that is based on ``batch-switching'' PI (Kalyanakrishnan \textit{et al.}~\cite{Kalyanakrishnan}).

The upper bounds listed above for MDPs also apply to DMDPs. However, a much stronger result has been shown when PI is  applied to DMDPs. Post and Ye~\cite{PostYe} demonstrate that the max-gain variant of the Simplex indeed terminates after a polynomial number of iterations on $n$-state, $k$-action DMDP. In this variant, the improvement set is $\{(\bar{s}, \bar{a})\}$, where $\bar{s}, \bar{a} \in \argmax_{(s, a) \in S \times A} \rho^{\pi}(s, a)$, with ties broken arbitrarily. As we shall see in Section~\ref{sec:DMDPs}, every DMDP $M$ induces a directed multigraph $G_{M}$, with each policy inducing a subgraph that is guaranteed to contain a directed cycle. Post and Ye establish that the max-gain Simplex algorithm registers a significant jump in the objective function when proceeding from $\pi$ to $\pi^{\prime}$ if some cycle induced by $\pi$ is \textit{not} induced by $\pi^{\prime}$. Moreover, such a break of a cycle must occur within a polynomial number of iterations, resulting in an overall upper bound of $O(n^5 k^2 \log^2 n)$ iterations for the max-gain Simplex algorithm. This upper bound has subsequently been improved by a factor of $n$ (Hansen \text{et al.}~\cite{HansenKaplanZwick}) and also generalised (Scherrer~\cite{Scherrer}). Even if the max-gain simplex algorithm is strongly polynomial for DMDPs, there do exist PI variants that are exponentially lower-bounded. Ashutosh \emph{et al.}~\cite{Ashutosh} construct an $n$-state, $k$-action whose PI-DAG has a path of length $\Omega(k^{n/2})$.

Unlike preceding analyses to obtain upper bounds for DMDPs (Post and Ye~\cite{PostYe}, Hansen \text{et al.}~\cite{HansenKaplanZwick}, Scherrer~\cite{Scherrer}), ours does not principally rely on geometric properties such as the objective function of policies. Rather, our arguments are primarily based on bounding discrete quantities: the number of directed cycles induced by policies in certain subgraphs of $G_{M}$. We derive two graph-theoretic results in the next section, and apply them to the analysis of PI on DMDPs in Section~\ref{sec:DMDPs}.

%% file: cycles.tex
\section{Number of cycles in digraphs}
\label{sec:cycles}

In this section, we present results on the maximum number of cycles in directed multigraphs with certain constraints on degree and edge multiplicity. Our main results are theorems \ref{thm:nk} and \ref{thm:cycle-bound}, which plug into the analysis of PI on DMDPs in Section~\ref{sec:DMDPs}.

The problem of bounding the number of cycles in graphs has a long history. Bounds have been established in terms of several basic graph parameters, including the number of edges, vertices, minimum/maximum/average degree, and cyclomatic number (Ahrens~\cite{Ahrens}, Arman and Tsaturian~\cite{Arman17}, Dvo\v{r}\'ak \emph{et al.}~\cite{Dvorak}, Entringer and Slater~\cite{Entringer}, Guichard~\cite{Guichard}, Volkmann~\cite{Volkmann}). Furthermore, restrictions of this problem to specific classes of graphs are also well studied: planar graphs (Aldred and Thomassen~\cite{Aldred08}, Alt \emph{et al.}~\cite{Alt}, Buchin \emph{et al.}~\cite{Buchin}, Dvo\v{r}\'ak~\cite{Dvorak}), graphs with forbidden subgraphs (Morrison \emph{et al.}~\cite{Morrison}), Hamiltonian graphs (Rautenbach and Stella~\cite{Rautenbach}, Shi~\cite{Shi}), triangle-free graphs (Arman \emph{et al.}~\cite{Arman}, Durocher \emph{et al.}~\cite{Durocher}), random graphs (Tak\'acs~\cite{Lajos}), bipartite graphs (Alt \emph{et al.}~\cite{Alt}), $k$-connected graphs (Knor~\cite{Knor}), grid graphs (Alt \emph{et al.}~\cite{Alt}), outerplanar and series--parallel graphs (Mier and Noy~\cite{deMier}), complement of a tree (Reid~\cite{Reid}, Zhou~\cite{Zhou}), $3$-connected cubic (Hamiltonian) graphs (AlBdaiwi~\cite{AlBdaiwi}, Aldred and Thomassen~\cite{Aldred97}), and $3$-colorable triangulated graphs (Alt \emph{et al.}~\cite{Alt}).

Analogous literature for digraphs is relatively sparse. Yoshua Perl~\cite{Perl} proved bounds for directed multigraphs with a fixed number of edges. Some restrictions to specific classes of digraphs have also been studied: digraphs with large girth (Allender~\cite{Allender}) and digraphs with restricted cycle lengths (Gerbner \emph{et al.}~\cite{Gerbner}). A few other studies have considered the very closely related question of bounding the number of paths between two given vertices in the digraph: (acyclic) simple digraphs with a given number of edges (Delivorias and Richter~\cite{DelivoriasRichter}, Perl~\cite{Perl}) and simple acyclic digraphs with a given number of vertices and edges (Golumbic and Perl~\cite{GolumbicPerl}).

We shall consider directed multigraphs with restrictions on degree and edge multiplicity. In the next section (in particular, lemmas~\ref{lem:num-path-cycles-nk} and \ref{lem:num-path-cycles}), we illustrate how the bounds on the number of cycles (that we prove later in this section) can be used to establish bounds on the number of paths and path-cycles (see Section~\ref{sec:DMDPs} for definition) in the considered digraphs. Before we prove the main results of this section, we provide some basic definitions and set up some notation.

\subsection{Definitions.}

We use the term \emph{digraph} to refer to directed graphs possibly containing multi-edges and self-loops. A digraph is said to be \emph{simple} if it does not contain self-loops or multi-edges. For any digraph $G$, we shall denote its set of vertices and edges by $V(G)$ and $E(G)$, respectively. By an edge $(u,v) \in E(G)$, we mean the multi-edge from $u$ to $v$ in $G$ unless otherwise specified, and denote its multiplicity by $\text{mult}(u,v)$. For $V_0 \subseteq V(G)$, we denote the induced subgraph of $G$ on $V_0$ by $G[V_0]$. Finally, we shall denote the digraph obtained by deleting an edge $(u,v)$ from $G$ by $G \setminus (u,v)$ and the digraph obtained by contracting (defined below) an edge $(u,v)$ in $G$ by $G / (u,v)$.

\begin{definition}
    For integers $n \ge 0, k \ge 2$, we define $\mathcal{G}_{\text{simple}}^{n, k}$ as the set of all digraphs with $n$ vertices, outdegree at most $k$, with the additional restriction that the digraph does not contain any multi-edge.
\end{definition}

\begin{definition}
    For integers $n \ge 0, k \ge 2$, we define $\mathcal{G}_{\text{multi}}^{n, k}$ as the set of all digraphs with $n$ vertices, outdegree exactly $k$, with the restriction that the multiplicity of edges connecting distinct vertices is at most $k-1$.
\end{definition}

Note that digraphs in $\mathcal{G}_{\text{simple}}^{n, k}$ and $\mathcal{G}_{\text{multi}}^{n, k}$ might contain self-loops. Also note that $\mathcal{G}_{\text{simple}}^{0, k} = \mathcal{G}_{\text{multi}}^{0, k}$ and both of these sets contain a single element, the empty digraph $(\phi, \phi)$.

\begin{definition}
    Given a digraph $G$, its \emph{skeleton} $\text{Skel}(G)$ is defined as the digraph obtained by replacing each edge in $G$ with the corresponding edge of multiplicity $1$.
\end{definition}

We remark that a digraph can be specified by its skeleton and the multiplicities of its edges.

\begin{definition}
    Let $G$ be a digraph, and $(u,v)$ be an edge with distinct end points $u$ and $v$. The digraph $G/(u,v)$ obtained by contracting the edge $(u,v)$ in $G$ is defined as
    \begin{align*}
        G/(u,v) = (&V(G)\setminus \{u,v\} \cup \{w\},\\
        & E(G) \cup \{(w,y): (u,y) \in E(G) \text{ or } (v,y) \in E(G) \text{ with } y \not\in \{u, v\}\} \\
        &\quad \quad \;\; \cup \{(x,w): (x,v) \in E(G) \text{ or } (x,u) \in E(G) \text{ with } x \not\in \{u, v\}\} \\
        &\quad \quad \;\; \cup \{(w,w):(v,u)\in E(G)\text{ or }(u,u)\in E(G)\text{ or }(v,v)\in E(G)\} \\
        &\quad \quad \;\; \setminus \{(x,y) \in E(G): x, y \in \{u,v\}\}).
    \end{align*}
\end{definition}

The multi-edge contraction operation (or edge contraction as defined above) on $(u,v)\in E(G)$ replaces $u$ and $v$ with a single vertex $w$ such that all edges incident to $u$ or $v$, other than the said multi-edge, are now incident to $w$.

\begin{definition}
    Let $G$ be a simple digraph. An edge $(u,v) \in E(G)$ is said to be \emph{in}-contractible if there does not exist any vertex $x \in V(G)$ distinct from $u$ and $v$ such that $(x,u), (x,v) \in E(G)$. Similarly, $(u,v)$ is said to be \emph{out}-contractible if there does not exist any vertex $x \in V(G)$ distinct from $u$ and $v$ such that $(u,x), (v,x) \in E(G)$. Finally, $(u,v)$ is said to be \emph{contractible} if it is both in-contractible and out-contractible.
\end{definition}

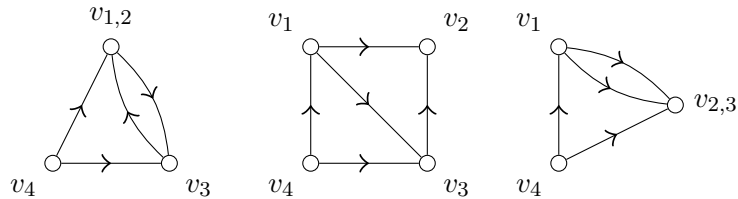
\begin{figure}[b]
\centering
\begin{subfigure}{.2\textwidth}

\[\begin{tikzpicture}[x=1.55cm, y=1.55cm,
    every edge/.style={
        draw,
        postaction={decorate,
                    decoration={markings,mark=at position 0.5 with {\arrow[scale=2]{>}}}
                   }
        },
    every loop/.style={},
    el/.style = {inner sep=2pt, align=left, sloped},
]

\vertex (v1v2) at  (0.5,1) [label={[label distance=0mm]90:$v_{1, 2}$}]{};
\vertex (v3) at  (1,0) [label={[label distance=0mm]315:$v_3$}]{};
\vertex (v4) at  (0,0) [label={[label distance=0mm]225:$v_4$}]{};

\path
    (v4) edge (v1v2)
    (v3) edge[bend left=20] (v1v2)
    (v4) edge (v3)
    (v1v2) edge[bend left=20] (v3)

;
\end{tikzpicture}\]
\end{subfigure}
\begin{subfigure}{.2\textwidth}

\[\begin{tikzpicture}[x=1.55cm, y=1.55cm,
    every edge/.style={
        draw,
        postaction={decorate,
                    decoration={markings,mark=at position 0.5 with {\arrow[scale=2]{>}}}
                   }
        },
    every loop/.style={},
    el/.style = {inner sep=2pt, align=left, sloped},
]

\vertex (v1) at  (0,1) [label={[label distance=0mm]135:$v_1$}]{};
\vertex (v2) at  (1,1) [label={[label distance=0mm]45:$v_2$}]{};
\vertex (v3) at  (1,0) [label={[label distance=0mm]315:$v_3$}]{};
\vertex (v4) at  (0,0) [label={[label distance=0mm]225:$v_4$}]{};

\path
    (v4) edge (v1)
    (v1) edge (v2)
    (v3) edge (v2)
    (v4) edge (v3)
    (v1) edge (v3)

;
\end{tikzpicture}\]
\end{subfigure}
\begin{subfigure}{.2\textwidth}

\[\begin{tikzpicture}[x=1.55cm, y=1.55cm,
    every edge/.style={
        draw,
        postaction={decorate,
                    decoration={markings,mark=at position 0.5 with {\arrow[scale=2]{>}}}
                   }
        },
    every loop/.style={},
    el/.style = {inner sep=2pt, align=left, sloped},
]

\vertex (v1) at  (0,1) [label={[label distance=0mm]135:$v_1$}]{};
\vertex (v2v3) at  (1,0.5) [label={[label distance=0mm]0:$v_{2, 3}$}]{};
\vertex (v4) at  (0,0) [label={[label distance=0mm]225:$v_4$}]{};

\path
    (v4) edge (v1)
    (v1) edge[bend left=20] (v2v3)
    (v4) edge (v2v3)
    (v1) edge[bend right=20] (v2v3)

;
\end{tikzpicture}\]
\end{subfigure}
\caption{An example of a simple digraph $G$ (centre). The edges $(v_1,v_2)$ and $(v_4,v_3)$ are contractible, $(v_4,v_1)$ is not out-contractible, $(v_3,v_2)$ is not in-contractible, and $(v_1,v_3)$ is neither in-contractible nor out-contractible. Contraction of the edge $(v_1, v_2)$ leads to the graph $G/(v_1,v_2)$ (left) with no multi-edges since $(v_1,v_2)$ is contractible, while contraction of the edge $(v_3, v_2)$ leads to the graph $G/(v_3,v_2)$ (right) containing a multi-edge since $(v_1,v_2)$ is not in-contractible.}
\label{fig:contraction}
\end{figure}

We refer the reader to Figure~\ref{fig:contraction} for an example illustrating the above notions of contractibility and the edge contraction operation.

It follows directly from the definition that if an edge $(u,v)$ in a simple digraph $G$ is contractible, it can be contracted without forming multi-edges in the resulting digraph. Note that the digraph obtained may contain self-loops even if the above property is satisfied.

We use the term \emph{cycle} to refer to a directed cycle unless otherwise specified. For a digraph $G$, we shall denote the number of cycles in $G$ by $C(G)$. Further, for a vertex $v \in V(G)$, we denote the number of cycles in $G$ passing through $v$ by $C(G,v)$. Similarly, for a multi-edge $e \in E(G)$, we denote the number of cycles in $G$ passing through $e$ by $C(G,e)$. Finally, for $E \subseteq E(G)$, we denote the number of cycles in $G$ passing through at least one edge in $E$ by $C(G,E)$. We remark that our definitions incorporate the multiplicity of edges while computing the number of cycles, i.e., cycles passing through distinct edges that are part of the same multi-edge are considered distinct.

Now, we define $M_k(n) = \max_{G \in \mathcal{G}_{\text{simple}}^{n, k}} C(G)$: that is, $M_k(n)$ denotes the maximum number of cycles in any digraph in $\mathcal{G}_{\text{simple}}^{n, k}$. Similarly, we define $F_k(n) = \max_{G \in \mathcal{G}_{\text{multi}}^{n, k}} C(G)$: that is, $F_k(n)$ denotes the maximum number of cycles in any digraph in $\mathcal{G}_{\text{multi}}^{n, k}$.

\begin{definition}
    Given digraphs $G$ and $H$, we say that $G$ is $H$-free if it has no subgraph isomorphic to $H$.
\end{definition}

\subsection{Bounds on the number of cycles.}

We now prove upper bounds on $M_k(n)$ and $F_k(n)$.

\subsubsection{Bounds for simple digraphs.}

Dvo\v{r}\'ak \emph{et al.}~\cite{Dvorak} proved a more general version of the theorem below for the case of undirected graphs. Their proof idea works for digraphs, as well. We include the full proof for the sake of completeness.

\begin{theorem}
\label{thm:nk}
    For integers $n \ge 0$, $k \ge 2$, $M_k(n) \le (k+1)!^{n/(k+1)}$.
\end{theorem}

\proof
    The result clearly holds for $n = 0, 1,$ and $2$ since $M_k(0) = 0, M_k(1) = 1,$ and $M_k(2) \le 3 \le (k+1)!^{2/(k+1)}$ for all $k \ge 2$. Further, it is easy to check that $M_2(3) = 5 \le 6 = (2+1)!^{3/(2+1)}$ and $M_k(3) = 8 \le (k+1)!^{3/(k+1)}$ for all $k \ge 3$.
    We shall henceforth assume that $n \ge 4$. Let $G \in \mathcal{G}_{\text{simple}}^{n, k}$ and $V(G) = \{v_1, v_2, \dots, v_n\}$. For $1 \le i \le n$, let $\ell_i \in \{0, 1\}$ denote the number of self-loops on $v_i$ in $G$ and $d_i = \text{outdegree}(v_i)$. Further, let $G_0$ be the simple digraph obtained by removing all self-loops from $G$. Let $A_0$ be the adjacency matrix of $G_0$ and $A' = A_0 + I$. For each cycle $C$ in $G_0$, there exists a unique permutation $\sigma \in \text{Sym}(n)$ whose cycle decomposition contains the cycle $C$ and fixes the rest of the vertices. Therefore, $C(G_0)$ is less than or equal to the permanent of $A'$, which is defined by
    \begin{equation*}
        \text{perm}(A') = \sum_{\sigma \in \text{Sym}(n)} \prod_{i = 1}^n a_{i, \sigma(i)}'.
    \end{equation*}
    Note that the sum of the entries in the $i$th row of $A'$ is equal to $d_i + 1 - \ell_i$. Hence, using Br\`egman's theorem (Br\`egman~\cite{Bregman}), we obtain
    \begin{equation*}
        \text{perm}(A') \le \prod_{i=1}^n (d_i + 1 - \ell_i)!^{1/(d_i + 1 - \ell_i)},
    \end{equation*}
    which further yields
    \begin{equation}
    \label{eqn:perm}
        C(G_0) \le \text{perm}(A') \le \prod_{i=1}^n (k + 1 - \ell_i)!^{1/(k + 1 - \ell_i)}
    \end{equation}
    since $d_i \le k$ for each $1 \le i \le n$ and the function $f: \mathbb{N} \to \mathbb{R}$ defined by $f(m) = m!^{1/m}$ for $m \in \mathbb{N}$ is monotonically increasing. Now, using $C(G) = C(G') + \sum_{i=1}^n \ell_i$ in \eqref{eqn:perm}, we obtain
    \begin{equation}
    \label{eqn:ineq-cycles}
        C(G) \le \prod_{i=1}^n (k + 1 - \ell_i)!^{1/(k + 1 - \ell_i)} + \sum_{i=1}^n \ell_i.
    \end{equation}
    We shall now show that $C(G) \le (k+1)!^{n/(k+1)}$. If $\ell_i = 0$ for all $1 \le i \le n$, then \eqref{eqn:ineq-cycles} yields $C(G) \le (k+1)!^{n/(k+1)}$. Therefore, we may assume $\ell_1 = 1$ without loss of generality. We show that changing the value of $\ell_1$ to $0$ while keeping $l_{2}, l_{3}, \dots, l_{n}$ fixed increases the value of the RHS of \eqref{eqn:ineq-cycles}. For $k \ge 2$, we have $(k+1)!^{1/(k+1)} - k!^{1/k} \ge 1/e$, and
    \begin{equation*}
        \prod_{i=2}^n (k + 1 - \ell_i)!^{1/(k + 1 - \ell_i)} \ge 2^{(n-1)/2} \ge 2^{3/2}.
    \end{equation*}
    Combining these two inequalities, we obtain
    \begin{equation*}
        k!^{1/k} \prod_{i=2}^n (k + 1 - \ell_i)!^{1/(k + 1 - \ell_i)} + \frac{2^{3/2}}{e} \le (k+1)!^{1/(k+1)} \prod_{i=2}^n (k + 1 - \ell_i)!^{1/(k + 1 - \ell_i)},
    \end{equation*}
    which further yields
    \begin{equation*}
        k!^{1/k} \prod_{i=2}^n (k + 1 - \ell_i)!^{1/(k + 1 - \ell_i)} + 1 + \sum_{i=2}^n \ell_i \le (k+1)!^{1/(k+1)} \prod_{i=2}^n (k + 1 - \ell_i)!^{1/(k + 1 - \ell_i)} + 0 + \sum_{i=2}^n \ell_i.
    \end{equation*}
    Repeating the same argument for each $\ell_i$ that is equal to $1$, we conclude that changing the value of $\ell_i$ to $0$ for all $1 \le i \le n$ increases the value of the RHS of \eqref{eqn:ineq-cycles}. Hence, $C(G) \le (k+1)!^{n/(k+1)}$.
    
    Since the above bound holds for each $G \in \mathcal{G}_{\text{simple}}^{n, k}$, we obtain the desired result.
\endproof

\begin{figure}[b]
\centering
\[\begin{tikzpicture}[x=1cm, y=1cm,
        every edge/.style={
            draw,
            postaction={decorate,
                        decoration={markings,mark=at position 0.41 with {\arrow[scale=2]{>}}}
                       }
            }
    ]
    
    \vertex (v00) at  (0,-2) [label=below:$v_{0,0}$]{};
    
    \vertex (v01) at  (-0.866,-0.5) [label=above right:$v_{0,1}$]{};
    \vertex (v02) at  (-2.598,-0.5) [label=left:$v_{0,2}$]{};
    \vertex (v03) at  (-1.732,-2) [label=below left:$v_{0,3}$]{};
    
    \vertex (v10) at  (-1.732,1) [label=above left:$v_{1,0}$]{};
    
    \vertex (v11) at  (0,1) [label=below:$v_{1,1}$]{};
    \vertex (v12) at  (0.866,2.5) [label=above:$v_{1,2}$]{};
    \vertex (v13) at  (-0.866,2.5) [label=above:$v_{1,3}$]{};
    
    \vertex (v20) at  (1.732,1) [label=above right:$v_{2,0}$]{};
    
    \vertex (v21) at  (0.866,-0.5) [label=above left:$v_{2,1}$]{};
    \vertex (v22) at  (1.732,-2) [label=below right:$v_{2,2}$]{};
    \vertex (v23) at  (2.598,-0.5) [label=right:$v_{2,3}$]{};
    
    \path
    (v00) edge (v01)
    (v00) edge (v02)
    (v00) edge (v03)
    
    (v01) edge[bend right=15] (v02)
    (v01) edge[bend right=15] (v03)
    (v02) edge[bend right=15] (v01)
    (v02) edge[bend right=15] (v03)
    (v03) edge[bend right=15] (v01)
    (v03) edge[bend right=15] (v02)
    
    (v01) edge (v10)
    (v02) edge (v10)
    (v03) edge (v10)
    
    (v10) edge (v11)
    (v10) edge (v12)
    (v10) edge (v13)
    
    (v11) edge[bend right=15] (v12)
    (v11) edge[bend right=15] (v13)
    (v12) edge[bend right=15] (v11)
    (v12) edge[bend right=15] (v13)
    (v13) edge[bend right=15] (v11)
    (v13) edge[bend right=15] (v12)
    
    (v11) edge (v20)
    (v12) edge (v20)
    (v13) edge (v20)
    
    (v20) edge (v21)
    (v20) edge (v22)
    (v20) edge (v23)
    
    (v21) edge[bend right=15] (v22)
    (v21) edge[bend right=15] (v23)
    (v22) edge[bend right=15] (v21)
    (v22) edge[bend right=15] (v23)
    (v23) edge[bend right=15] (v21)
    (v23) edge[bend right=15] (v22)
    
    (v21) edge (v00)
    (v22) edge (v00)
    (v23) edge (v00)
    ;
    \end{tikzpicture}\]

\caption{The digraph $G_\text{example}^{3,3}$.}
\label{fig:3G3}
\end{figure}
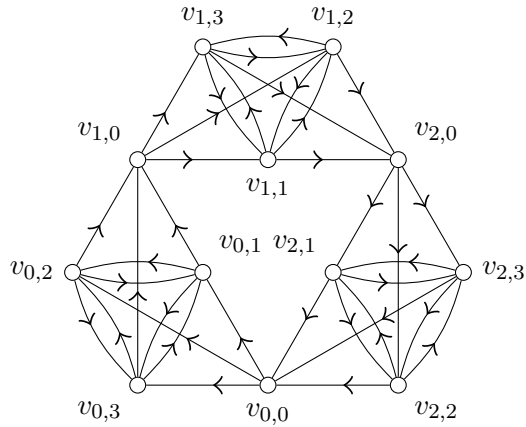

We now provide a particular example of a digraph in $\mathcal{G}_{\text{simple}}^{n, k}$. For $k \ge 2$, $\ell \ge 1$, we define the simple digraph $G_\text{example}^{\ell,k}$ which consists of $\ell$ units of single vertices and $k$-cliques arranged in an alternating cyclic fashion. Formally, $G_\text{example}^{\ell,k} = (V, E)$, where $V = \{v_{i,j}: 0 \le i < \ell, 0 \le j \le k\}$ and $E = \{(v_{i,0}, v_{i,j}): 0 \le i < \ell, 1 \le j \le k\} \cup \{(v_{i,j_1}, v_{i,j_2}): 0 \le i < \ell, 1 \le j_1, j_2 \le k\} \cup \{(v_{i_1,j}, v_{i_2,0}): i_2 - i_1 \equiv 1 \text{ mod }n, 0 \le i_1 \le \ell, 0 \le i_2 \le \ell, 1 \le j \le k\}$. See Figure~\ref{fig:3G3} for an example digraph $G_\text{example}^{3,3}$. It can be shown using direct enumeration that
\begin{equation}
    C(G_\text{example}^{\ell,k}) = \ell \left(2^{k+1} - \binom{k}{2} - 2k - 2\right) + \left(\sum_{r=0}^{k} \frac{k!}{r!} - 1\right)^\ell.
\end{equation}
We remark that the upper bound on $M_k(n)$ in Theorem~\ref{thm:nk} is asymptotically sharp in the sense that
\begin{equation}
    \lim_{k \to \infty} \left(\frac{C(G_\text{example}^{\ell,k})}{(k+1)!^\ell}\right)^{1/(\ell(k+1))} = 1,
\end{equation}
which implies that $M_k(n)^{1/n}$ is equal to $(k+1)!^{1/(k+1)}$ upto a multiplicative factor that gets arbitrarily close to $1$ as $k$ goes to infinity.

\subsubsection{Bounds for directed multigraphs.}

We begin by providing two particular examples of digraphs in $\mathcal{G}_{\text{multi}}^{n, k}$, with the same skeleton. For $n \ge 3$, we define the simple digraph $G_n = (V,E)$ with $V = \{v_i: 0 \le i < n\}$ and $E = \{(v_i,v_j): j-i \equiv 1\text{ mod }n\text{ or }j-i \equiv 2\text{ mod }n\}$.
Let $G_{n,k}$ be the digraph with skeleton $\text{Skel}(G_{n,k}) = G_n$ in which edges $(v_i,v_j)$ with $j-i \equiv 1\text{ mod }n$ have multiplicity $k-1$ while the remaining edges have multiplicity $1$.
Similarly, let $G'_{n,k}$ be the digraph with skeleton $\text{Skel}(G'_{n,k}) = G_n$ in which edges $(v_i,v_j)$ with $j-i \equiv 1\text{ mod }n$ having multiplicity equal to $1$ while the remaining edges have multiplicity $k-1$.
Note that $G_{n,2}$, $G'_{n,2}$ and $G_n$ are isomorphic. From the above definitions, it is easy to see that $G_{n,k}, G'_{n,k} \in \mathcal{G}_{\text{multi}}^{n, k}$.
Further, it can be shown using direct enumeration that
\begin{equation}
\label{eqn:cycles-recurrence-1}
    C(G_{n,k}) =
    \begin{cases} 
        S_{n-2} + S_{n} + 1, & \text{if }n\text{ is odd,} \\
        S_{n-2} + S_{n}, & \text{otherwise},
    \end{cases}
\end{equation}
where $S_n$ is defined by the recurrence relation $S_n = (k-1)S_{n-1} + S_{n-2}$ with boundary condition $S_0 = 1, S_1 = k-1$. For any vertex $v \in V(G_{n,k})$, $C(G_{n,k})$ can be written as a sum of the number of cycles passing through $v$ and those not passing through $v$. In this case, $C(G_{n,k}) - C(G_{n,k}, v) = S_{n-2}$ and the $1$ extra cycle contributing to $C(G_{n,k}, v)$ for odd $n$ corresponds to the Hamiltonian cycle comprised of all edges of multiplicity $1$ in $G_{n,k}$.
Similarly, one can also show that
\begin{equation}
\label{eqn:cycles-recurrence-2}
    C(G'_{n,k}) =
    \begin{cases} 
        (k-1)T_{n-2} + T_{n} + (k-1)^n, & \text{if }n\text{ is odd,} \\
        (k-1)T_{n-2} + T_{n}, & \text{otherwise},
    \end{cases}
\end{equation}
where $T_n$ is defined by the recurrence relation $T_n = T_{n-1} + (k-1)T_{n-2}$ with boundary condition $T_0 = 1, T_1 = 1$.

In Lemma~\ref{lem:order-cycles}, we show that the number of cycles in $G_{n,k}$ is greater than or equal to the number of cycles in $G'_{n,k}$.

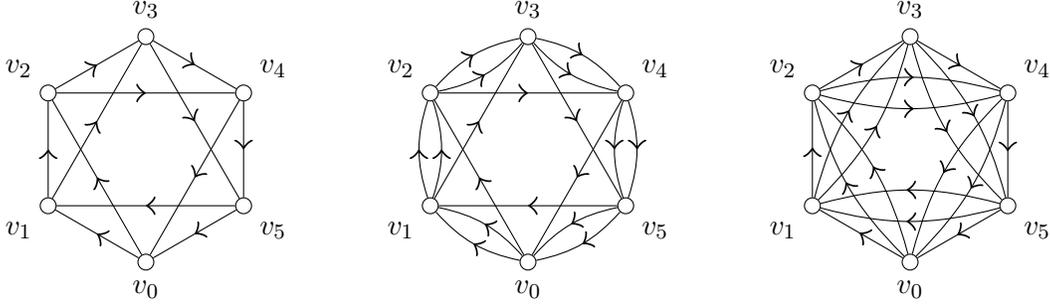
\begin{figure}[!t]
    \centering
    \begin{subfigure}{.3\textwidth}
    \[\begin{tikzpicture}[x=0.75cm, y=0.75cm,
        every edge/.style={
            draw,
            postaction={decorate,
                        decoration={markings,mark=at position 0.5 with {\arrow[scale=2]{>}}}
                       }
            }
    ]
    
    \vertex (v0) at  (0,-2) [label=below:$v_{0}$]{};
    \vertex (v1) at  (-1.732,-1) [label=below left:$v_{1}$]{};
    \vertex (v2) at  (-1.732,1) [label=above left:$v_{2}$]{};
    \vertex (v3) at  (0,2) [label=above:$v_{3}$]{};
    \vertex (v4) at  (1.732,1) [label=above right:$v_{4}$]{};
    \vertex (v5) at  (1.732,-1) [label=below right:$v_{5}$]{};
    
    \path
        (v0) edge (v1)
        (v1) edge (v2)
    	(v2) edge (v3)
    	(v3) edge (v4)
    	(v4) edge (v5)
    	(v5) edge (v0)
    	
    	(v0) edge (v2)
        (v1) edge (v3)
    	(v2) edge (v4)
    	(v3) edge (v5)
    	(v4) edge (v0)
    	(v5) edge (v1)
    ;
    \end{tikzpicture}\]
    \end{subfigure}
    \begin{subfigure}{.3\textwidth}
    \[\begin{tikzpicture}[x=0.75cm, y=0.75cm,
        every edge/.style={
            draw,
            postaction={decorate,
                        decoration={markings,mark=at position 0.5 with {\arrow[scale=2]{>}}}
                       }
            }
    ]
    
    \vertex (v0) at  (0,-2) [label=below:$v_{0}$]{};
    \vertex (v1) at  (-1.732,-1) [label=below left:$v_{1}$]{};
    \vertex (v2) at  (-1.732,1) [label=above left:$v_{2}$]{};
    \vertex (v3) at  (0,2) [label=above:$v_{3}$]{};
    \vertex (v4) at  (1.732,1) [label=above right:$v_{4}$]{};
    \vertex (v5) at  (1.732,-1) [label=below right:$v_{5}$]{};
    
    \path
        (v0) edge[bend right=18] (v1)
        (v1) edge[bend right=18] (v2)
    	(v2) edge[bend right=18] (v3)
    	(v3) edge[bend right=18] (v4)
    	(v4) edge[bend right=18] (v5)
    	(v5) edge[bend right=18] (v0)
    	
    	(v0) edge[bend left=18] (v1)
        (v1) edge[bend left=18] (v2)
    	(v2) edge[bend left=18] (v3)
    	(v3) edge[bend left=18] (v4)
    	(v4) edge[bend left=18] (v5)
    	(v5) edge[bend left=18] (v0)
    	
    	(v0) edge (v2)
        (v1) edge (v3)
    	(v2) edge (v4)
    	(v3) edge (v5)
    	(v4) edge (v0)
    	(v5) edge (v1)
    ;
    \end{tikzpicture}\]
    \end{subfigure}
    \begin{subfigure}{.3\textwidth}
    \[\begin{tikzpicture}[x=0.75cm, y=0.75cm,
        every edge/.style={
            draw,
            postaction={decorate,
                        decoration={markings,mark=at position 0.52 with {\arrow[scale=2]{>}}}
                       }
            }
    ]
    
    \vertex (v0) at  (0,-2) [label=below:$v_{0}$]{};
    \vertex (v1) at  (-1.732,-1) [label=below left:$v_{1}$]{};
    \vertex (v2) at  (-1.732,1) [label=above left:$v_{2}$]{};
    \vertex (v3) at  (0,2) [label=above:$v_{3}$]{};
    \vertex (v4) at  (1.732,1) [label=above right:$v_{4}$]{};
    \vertex (v5) at  (1.732,-1) [label=below right:$v_{5}$]{};
    
    \path
        (v0) edge (v1)
        (v1) edge (v2)
    	(v2) edge (v3)
    	(v3) edge (v4)
    	(v4) edge (v5)
    	(v5) edge (v0)
    	
    	(v0) edge[bend right=15] (v2)
        (v1) edge[bend right=15] (v3)
    	(v2) edge[bend right=15] (v4)
    	(v3) edge[bend right=15] (v5)
    	(v4) edge[bend right=15] (v0)
    	(v5) edge[bend right=15] (v1)
    	
    	(v0) edge[bend left=15] (v2)
        (v1) edge[bend left=15] (v3)
    	(v2) edge[bend left=15] (v4)
    	(v3) edge[bend left=15] (v5)
    	(v4) edge[bend left=15] (v0)
    	(v5) edge[bend left=15] (v1)
    ;
    \end{tikzpicture}\]
    \end{subfigure}
    \caption{The digraphs $G_6$, $G_{6,3}$ and $G'_{6,3}$ (from left to right).}
    \label{fig:digraph-examples}
\end{figure}

\begin{lemma}
\label{lem:order-cycles}
    For natural numbers $n \ge 3$, $k \ge 2$, $C(G_{n,k}) \ge C(G'_{n,k})$.
\end{lemma}

\proof
    We will show using induction that $(k-1)T_{n-2} + T_n + (k-1)^n \le S_{n-2} + S_n + 1$ for all $n \ge 3$, thereby implying $C(G_{n,k}) \ge C(G'_{n,k})$ for each $n \ge 3$. For simplicity, let $L_n$ and $R_n$ denote the LHS and RHS of the above inequality, respectively. It is easy to check that $L_3 = R_3$ and $L_4 + 2(k-2)^2 = R_4$. Now, by induction hypothesis, we have $L_{n-1} \le R_{n-1}$ and $L_n \le R_n$, which implies $(k-1)L_n + L_{n-1} \le (k-1)R_n + R_{n-1}$, which when expanded and rearranged yields
    \begin{equation}
    \label{eqn:induction-inequality}
        (k-1)((k-1)T_{n-2} + T_{n-3}) + ((k-1)T_n + T_{n-1}) + (k-1)^{n+1} + (k-1)^{n-1} \le S_{n-1} + S_{n+1} + k.
    \end{equation}
    Since $\{T_n\}$ is a monotonically increasing sequence, $(k-1)T_{m} + T_{m-1} \ge T_m + (k-1) T_{m-1} = T_{m+1}$ for all $m \in \mathbb{N}$. Using this inequality and $(k-1)^{n-1} \ge k-1$ in \eqref{eqn:induction-inequality}, we obtain
    \begin{equation*}
        (k-1) T_{n-1} + T_{n+1} + (k-1)^{n+1}\le S_{n-1} + S_{n+1} + 1.
    \end{equation*}
\endproof

We now compute $C(G_{n,k})$. Solving the recurrence relation for $S_n$ with appropriate boundary conditions, we obtain
\begin{equation*}
    S_n = \frac{\left(\frac{k-1 + \sqrt{(k-1)^2 + 4}}{2}\right)^{n+1} - \left(\frac{k-1 - \sqrt{(k-1)^2 + 4}}{2}\right)^{n+1}}{\sqrt{(k-1)^2 + 4}},
\end{equation*}
which when used in \eqref{eqn:cycles-recurrence-1} yields $C(G_{n,k}) = \left\lceil \alpha(k)^n \right\rceil$, where $\lceil . \rceil$ is the ceiling function and
\begin{equation}
\label{eqn:alpha-k}
    \alpha(k) = \frac{k-1 + \sqrt{(k-1)^2 + 4}}{2}.
\end{equation}

\begin{remark}
    The digraph $G_n$ is an example of a Cayley graph (Meier~\cite[see Section~1.5]{Meier2008}) since $G_n = (V, \{(v,v+s): v \in V, s \in S\})$, where $V = \mathbb{Z}/N\mathbb{Z}$ and $S = \{1,2\}$ is a generating set for $G$. Cayley graphs are known to be extremal examples for various problems in graph theory.
\end{remark}

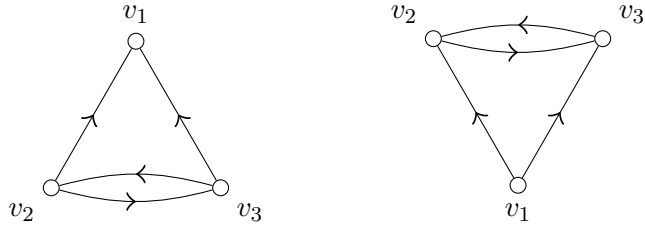
\begin{figure}[b]
    \centering
    \begin{subfigure}{.3\textwidth}
    \[\begin{tikzpicture}[x=0.65cm, y=0.65cm,
        every edge/.style={
            draw,
            postaction={decorate,
                        decoration={markings,mark=at position 0.5 with {\arrow[scale=2]{>}}}
                       }
            }
    ]
    
    \vertex (v1) at  (0,1) [label=above:$v_{1}$]{};
    \vertex (v2) at  (-1.732,-2) [label=below left:$v_{2}$]{};
    \vertex (v3) at  (1.732,-2) [label=below right:$v_{3}$]{};
    
    \path
        (v2) edge[bend right=15] (v3)
        (v3) edge[bend right=15] (v2)
    	
    	(v2) edge (v1)
        (v3) edge (v1)
    ;
    \end{tikzpicture}\]
    \end{subfigure}
    \begin{subfigure}{.3\textwidth}
    \[\begin{tikzpicture}[x=0.65cm, y=0.65cm,
        every edge/.style={
            draw,
            postaction={decorate,
                        decoration={markings,mark=at position 0.5 with {\arrow[scale=2]{>}}}
                       }
            }
    ]
    
    \vertex (v1) at  (0,-2) [label=below:$v_{1}$]{};
    \vertex (v2) at  (-1.732,1) [label=above left:$v_{2}$]{};
    \vertex (v3) at  (1.732,1) [label=above right:$v_{3}$]{};
    
    \path
        (v2) edge[bend right=15] (v3)
        (v3) edge[bend right=15] (v2)
    	
    	(v1) edge (v2)
        (v1) edge (v3)
    ;
    \end{tikzpicture}\]
    \end{subfigure}
    \caption{The digraphs $H_1$ and $H_2$ (from left to right).}
    \label{fig:h1h2}
\end{figure}

For $i \in \{1,2\}$, let $H_i$ be the digraph $(V,E_i)$, where $V = \{v_1, v_2, v_3\}$, $E_1 = \{(v_2,v_1), (v_3,v_1), (v_2,v_3),\\(v_3,v_2)\}$, and $E_2 = \{(v_1,v_2), (v_1,v_3), (v_2,v_3), (v_3,v_2)\}$. See Figure~\ref{fig:h1h2} for drawings of $H_1$ and $H_2$.

In Lemma~\ref{lem:structure}, we prove a bound on the number of edges that are not in-contractible in a simple two-regular $H_1$-free and $H_2$-free digraph. Furthermore, we characterise the graphs for which this bound is achieved. In Lemma~\ref{lem:bound}, we provide an upper bound on the number of cycles for these characterised graphs. These results are used in the proof of Theorem~\ref{thm:cycle-bound} in the following way: if the graph has few edges that are not in-contractible, then we contract edges to obtain a recursive bound on the number of cycles; if the graph has many edges that are not in-contractible, then we use Lemma~\ref{lem:bound} to bound the number of cycles.


\begin{lemma}
\label{lem:structure}
    Let $G$ be a simple two-regular $H_1$-free and $H_2$-free digraph on $n \ge 4$ vertices. Then, $G$ can have at most $n$ edges that are not in-contractible. Further, $G$ has exactly $n$ edges that are not in-contractible if and only if each connected component of $G$ is isomorphic to $G_m$ for some $m \ge 4$.
\end{lemma}

\proof
    For any $v \in V(G)$, we have distinct vertices $x, y \in V(G)$ such that $(x,v), (y,v) \in E(G)$ since $G$ is simple.
    If the edge $(x,v)$ is not in-contractible, then $(y,x) \in E(G)$. Similarly, if the edge $(y,v)$ is not in-contractible, then $(x,y) \in E(G)$. Suppose both $(x,v)$ and $(y,v)$ are not in-contractible. Then, the subgraph $(\{x,y,v\},\{(x,v),(y,v),(x,y),(y,x)\})$ of $G$ is isomorphic to $H_1$, a contradiction. The sets consisting of incoming edges to a vertex in $G$ constitute a uniform $n$-partition of $E(G)$. Since at most one edge in each such set is not in-contractible, the digraph $G$ can have at most $n$ edges that are not in-contractible.
    

    Now, suppose that $G$ has exactly $n$ edges that are not in-contractible. Then, for any vertex $v$ in $G$, exactly one of the two incoming edges to $v$ must be in-contractible. Let us pick a vertex $v_1 \in G$ with $(v_2,v_1),(v_3,v_1) \in E(G)$. Without loss of generality, we may assume that the edge $(v_2,v_1)$ is not in-contractible: that is, $(v_3,v_2) \in E(G)$. The other incoming edge to $v_2$ cannot originate from (i) $v_1$ since otherwise $G$ would contain a subgraph isomorphic to $H_2$, (ii) $v_2$ since otherwise $G$ would contain a self-loop, and (iii) $v_3$ since otherwise $G$ would contain a multi-edge. Therefore, it originates from a vertex $v_4 \not \in \{v_1,v_2,v_3\}$. Now, among the incoming edges $(v_3,v_2)$ and $(v_4,v_2)$ to $v_2$, the edge $(v_4,v_2)$ must be in-contractible since otherwise $(v_3,v_4) \in E(G)$, which is a contradiction to the fact that $\text{outdegree}(v_3) = 2$. Therefore, the edge $(v_3,v_2)$ is not in-contractible and hence $(v_4,v_3) \in E(G)$. The other incoming edge to $v_3$ cannot originate from (i) $v_2$ since otherwise $G$ would contain a subgraph isomorphic to $H_1$, (ii) $v_3$ since otherwise $G$ would contain a self-loop, and (iii) $v_4$ since otherwise $G$ would contain a multi-edge. Therefore, the other incoming edge to $v_3$ could either originate from $v_1$ or a vertex $v_5 \not \in \{v_1,v_2,v_3,v_4\}$.
    
    In the case where the other incoming edge to $v_3$ originates from $v_1$, the edge $(v_1,v_3)$ must be in-contractible since otherwise $(v_4,v_1) \in E(G)$, which is a contradiction to the fact that $\text{outdegree}(v_4) = 2$. Therefore, the edge $(v_4,v_3)$ is not in-contractible and hence $(v_1,v_4) \in E(G)$. Finally, $(v_2,v_4) \in E(G)$ since one incoming edge to $v_4$ must not be in-contractible, yielding a connected component $G_4$ of $G$.
    
    In the case where the other incoming edge to $v_3$ originates from a vertex $v_5 \not \in \{v_1,v_2,v_3,v_4\}$, we have $(v_5,v_4) \in E(G)$. Now, the other incoming edge to $v_4$ could either originate from $v_1,v_2$ or a vertex $v_6 \not \in \{v_1,v_2,v_3,v_4,v_5\}$ since the outdegree of $v_3, v_4,$ and $v_5$ is already satisfied. It cannot originate from $v_2$ since both incoming edges to $v_4$ would otherwise be in-contractible. If it originates from $v_1$, we get a connected component $G_5$ of $G$. If it originates from a vertex $v_6 \not \in \{v_1,v_2,v_3,v_4,v_5\}$, we continue in a similar way until we get a connected component $G_m$ of $G$ for some $m > 5$.
    
    Conversely, suppose each connected component of $G$ is isomorphic to $G_m$ for some $m \ge 4$. Then, within a connected component $G_m$, it is easy to check that the $m$ edges $(v_i,v_j)$ with $j-i \equiv 1\text{ mod }m$ are not in-contractible while the $m$ edges with $j-i \equiv 2\text{ mod }m$ are in-contractible. Summing over the connected components of $G$, we get exactly $n$ edges that are not in-contractible in $G$.
\endproof

\begin{remark}
    Under the same hypothesis as Lemma~\ref{lem:structure}, one can prove the following analogous symmetric result (although we require only one of these results): $G$ can have at most $n$ edges that are not out-contractible. Further, $G$ has exactly $n$ edges that are not out-contractible if and only if each connected component of $G$ is isomorphic to $G_m$ for some $m \ge 4$.
\end{remark}

\begin{lemma}
\label{lem:bound}
    Let $G$ be a digraph with $n \ge 4$ vertices, each of whose connected components is $G_{m,k}$ or $G'_{m,k}$ for some $m \ge 4$. Then, $C(G) \le (k-1)F_k(n-1) + F_k(n-2)$.
\end{lemma}

\proof
    For any natural number $m \ge 3$, we have $C(G_{m,k}) \le F_k(m)$ since $G_{m,k} \in \mathcal{G}_{\text{multi}}^{m, k}$. Using \eqref{eqn:cycles-recurrence-1} in this inequality, we obtain $S_{m-2} + S_{m} + 1 \le (k-1)F_k(m-1) + F_k(m-2)$.
    For natural numbers $p, q \ge 3$, we have
    \begin{align*}
        C(G_{p,k}) + C(G_{q,k}) &\le (S_{p-2} + S_{p} + 1) + (S_{q-2} + S_{q} + 1)\\
        &= (S_{p-2} + (S_{q-2} + 1)) + (S_{p} + (S_{q} + 1))\\
        &\le (S_{p+q-3} + S_{p+q-4}) + (S_{p+q-1} + S_{p+q-2})\\
        &\le S_{p+q-2} + S_{p+q}\\
        &\le C(G_{p+q,k}).
    \end{align*}
    Now, we have $C(G) = \sum_{i=1}^l C(G^i)$, where $G^i$ is the $i$-th connected component of $G$ and $l$ is the number of connected components in $G$.
    Let $m_i = |V(G^i)|$, so that $\sum_{i=1}^l m_i = n$.
    Using the subadditivity of the function $C(G_{.,k})$ and the fact that $C(G'_{m,k}) \le C(G_{m,k})$ for all $m \ge 3$, we get
    \begin{align*}
        C(G) &= \sum_{i=1}^l C(G^i)\\
        &\le \sum_{i=1}^l C(G_{m_i,k})\\
        &\le C(G_{\sum_{i=1}^l m_i,k})\\
        &= C(G_{n,k})\\
        &\le S_{n-2} + S_{n} + 1\\
        &\le (k-1)F_k(n-1) + F_k(n-2).
    \end{align*}
\endproof

We now prove the main result of this section, an asymptotically tight formula for $F_k(n)$. The proof proceeds by reducing the digraph into smaller digraphs belonging to the same family to obtain several non-homogeneous linear recursive bounds of order up to $3$ based on a case analysis, and finally combining all these bounds to obtain the final result.

\begin{theorem}
\label{thm:cycle-bound}
    For $k \ge 2$,
    \begin{equation*}
        F_k(n) = \Theta(\alpha(k)^n)
    \end{equation*}
    as $n \to \infty$, where $\alpha(k)$ is as defined in \eqref{eqn:alpha-k}. In particular, $F_2(n) = \Theta(F_n)$, where $F_n$ denotes the $n$-th Fibonacci number.
\end{theorem}

\proof
    We begin by making a few elementary observations about the function $F_k$. Clearly, $F_k(0) = 0, F_k(1) = k$, and $F_k(2) = \max(\{2k, (k-1)^2 + 2\})$, corresponding to the empty digraph, the digraph with a single vertex having a self-loop of multiplicity $k$, and the digraph with $2$ vertices, each having a self-loop of multiplicity $k$ or each having a self-loop of multiplicity $1$ along with an edge of multiplicity $k-1$ to the other vertex, respectively. For $n \in \mathbb{N}$, let $G^* \in \mathcal{G}_{\text{multi}}^{n-1, k}$ be such that $C(G^*) = F_k(n-1)$ (such a digraph exists since $\mathcal{G}_{\text{multi}}^{n-1, k}$ is finite) and let $G$ be the digraph obtained by adding a single vertex with a self-loop of multiplicity $k$ to $G^*$. Then, $F_k(n) \ge C(G) = C(G^*) + k = F_k(n-1) + k$. In particular, $F_k(n) \ge k$ for all $n \in \mathbb{N}$. Since we have already computed the values of $F_k(n)$ for $n \in \{0,1,2\}$, we shall henceforth assume that $n \ge 3$.
    
    Let $G \in \mathcal{G}_{\text{multi}}^{n, k}$. In the next three paragraphs of this proof, we argue that one may assume without loss of generality certain restrictions on the structure of $G$ without reducing the number of cycles or $C(G) \le (k-1) F_k(n-1) + F_k(n-2)$.
    
    Suppose that $G$ contains a vertex $v$ with a self-loop of multiplicity $k$. Then $C(G) = C(G[V(G) \setminus \{v\}]) + k \le F_k(n-1) + k \le (k-1) F_k(n-1) + F_k(n-2)$. Note that the digraph $G[V(G) \setminus \{v\}]$ might not necessarily be in $\mathcal{G}_{\text{multi}}^{n-1, k}$. However, one can add a self-loop of sufficient multiplicity to each vertex of $G[V(G) \setminus \{v\}]$ to obtain a digraph $G' \in \mathcal{G}_{\text{multi}}^{n-1, k}$ so that $C(G) \le C(G') \le F_k(n-1)$.
    
    Henceforth, we shall assume that each vertex in $G$ has at least one outgoing edge, which is not a self-loop.
    Note that for any vertex $v \in V(G)$, $C(G)$ is equal to the sum of the number of cycles passing through the vertex $v$ and those not passing through $v$. Let $e_1, e_2, \dots, e_k$ be the outgoing simple edges from $v$ (some of these edges might have the same end points since $G$ is a multigraph). Then, the number of cycles passing through $v$ is equal to the sum of the number of cycles passing through each of these edges. Now, there exists a permutation $\sigma \in \text{Sym}(k)$ such that $C(G, e_{\sigma(1)}) \ge C(G, e_{\sigma(2)}) \ge \dots \ge C(G, e_{\sigma(k)})$ and the end points of $e_{\sigma(1)}$ are not the same as the end points of $e_{\sigma(k)}$ (it is possible to satisfy the latter condition since $G$ does not contain edges of multiplicity $k$ connecting distinct vertices). In such a case, we can construct a digraph $G^*$ from $G$ by deleting $e_{\sigma(2)}, \dots, e_{\sigma(k-1)}$ and adding $(k-2)$ copies of $e_{\sigma(1)}$. The resulting digraph $G^*$ has the property $C(G^*) \ge C(G)$ and that the vertex $v$ in $G^*$ has two outgoing edges, one with multiplicity $1$ and the other with $k-1$.
    Applying this operation successively to every vertex in the digraph, we obtain a digraph $G'$ with $C(G) \le C(G')$, which also has the property that every vertex in $G'$ has two outgoing edges, one with multiplicity $1$ and the other with $k-1$. Further, for any vertex $v \in V(G')$, the number of cycles in $G'$ passing through the outgoing edge of multiplicity $k-1$ from $v$ is greater than $(k-1)$ times the number of cycles passing through the outgoing edge of multiplicity $1$ from $v$. Therefore, we may assume without loss of generality (renaming $G'$ to $G$) that each vertex in $\text{Skel}(G)$ has outdegree $2$ (one of these outgoing edges has multiplicity $k-1$ and the other has multiplicity $1$ in $G$).

    If $\text{Skel}(G)$ contains a vertex $v$ with a self-loop but no other incoming edge, we have $C(G) \le C(G[V(G) \setminus \{v\}]) + k-1 \le F_k(n-1) + k-1 \le (k-1)F_k(n-1) + F_k(n-2)$. If $\text{Skel}(G)$ contains an indegree $2$ vertex $v$ with a self-loop and the other incoming edge $(u,v)$ participating in a $2$-cycle, we have $C(G) \le C(G[V(G) \setminus \{v\}]) + (k-1)^2 + 1 \le F_k(n-1) + (k-1)k \le F_k(n-1) + (k-1)F_k(n-2) \le (k-1)F_k(n-1) + F_k(n-2)$.
    Finally, we consider the case where each vertex with a self-loop in $\text{Skel}(G)$ has an incoming edge other than the self-loop not participating in a $2$-cycle. For any such vertex $v$ and an incoming edge $(u,v)$ not participating in a $2$-cycle, the digraph $G'$ formed by deleting the self-loop on $v$ from $G$ and adding the edge $(v,u)$ with the same multiplicity as the self-loop, has at least as many cycles as the original digraph $G$. We repeatedly apply this operation to the digraph $G$ until no self-loops remain to obtain a digraph $G'$ satisfying $C(G) \le C(G')$. Therefore, we may assume without loss of generality (renaming $G'$ to $G$) that $G$ contains no self-loops.
    
    We now consider the following mutually exclusive exhaustive set of cases:

    \vspace{3pt}
    \noindent\textit{Case 1.} There is a vertex $v \in \text{Skel}(G)$ with $\text{indegree}(v) = 0$. In this case, we have $C(G) = C(G[V(G) \setminus \{v\}]) \le F_k(n-1) \le (k-1)F_k(n-1) + F_k(n-2)$ since none of the cycles in $G$ pass through $v$.

    \vspace{3pt}
    \noindent\textit{Case 2.} There is a vertex $v \in \text{Skel}(G)$ with $\text{indegree}(v) = 1$. Let $(u,v)$ be the unique incoming edge incident to $v$ in $G$. Let $(u,w)$ be the other outgoing edge from $u$ in $G$. The number of cycles passing through $(u,w)$ in $G$ is bounded above by the number of cycles in the digraph obtained by deleting the vertex $v$ and the incoming edges to $w$ other than the edge $(u,w)$, and contracting the edge $(u,w)$, which is less than or equal to $\text{mult}(u,w) F_k(n-2)$. Similarly, the number of cycles not passing through $(u,w)$ in $G$ is equal to the number of cycles in the digraph obtained by deleting the edge $(u,w)$ from $G$ and contracting the edge $(u,v)$, which is bounded above by $\text{mult}(u,v) F_k(n-1)$. Therefore, we obtain $F_k(n) \le \max(\{(k-1) F_k(n-1) + F_k(n-2), (k-1) F_k(n-2) + F_k(n-1)\}) = (k-1) F_k(n-1) + F_k(n-2)$.

    \vspace{3pt}
    \noindent\textit{Case 3.} All vertices in $\text{Skel}(G)$ have indegree equal to $2$: that is, $\text{Skel}(G)$ is $2$-regular.

    \vspace{2pt}
    \noindent\textit{Case 3.1.} Suppose that $k \ge 3$. We consider the case when $G$ contains a vertex $v$, both of whose incoming edges $(a,v)$ and $(b,v)$ have multiplicity $1$. We shall assume without loss of generality that $C(G,(b,v)) \ge C(G,(a,v))$. Let $(b,w)$ be the other outgoing edge from $b$ in $G$. Then, $C(G,(b,w)) \ge (k-1) C(G,(b,v))$. Further, we have
    \begin{equation*}
        C(G,v) = C(G,(a,v)) + C(G,(b,v)) \le 2 C(G,(b,v)),
    \end{equation*}
    and
    \begin{equation*}
        C(G) \ge C(G,b) = C(G,(b,v)) + C(G,(b,w)) \ge k C(G,(b,v)).
    \end{equation*}
    Combining the above inequalities, we obtain $C(G,v) \le 2C(G)/k$, which further implies $C(G) \le kC(G[V(G) \setminus \{v\}])/(k-2) \le k F_k(n-1)/(k-2)$. For $k \ge 4$, we have $k/(k-2) \le k-1$, which implies $C(G) \le (k-1) F_k(n-1)$.
    Now, we shall focus on the case $k = 3$. Let $(v,y)$ and $(v,z)$ be the outgoing edges from $v$ in $G$ with multiplicities $1$ and $2$, respectively. For $v_1 \in \{a,b\}$ and  $v_2 \in \{y,z\}$, we define $x_{v_1,v_2}$ to be the fraction of cycles in $G$ passing through the path $v_1 \rightarrow v \rightarrow v_2$; that is,
    \begin{equation*}
        x_{v_1,v_2} = C(G,\{(v_1,v),(v,v_2)\})/C(G).
    \end{equation*}
    Let $\boldsymbol{x} = [x_{a,y}\; x_{a,z}\; x_{b,y}\; x_{b,z}]^T$. We consider the following mutually exclusive exhaustive set of cases.

    \vspace{1pt}
    \noindent\textit{Case 3.1.1.} We first consider the case when $(a,z), (b,z) \not \in E(G)$. Let $G'$ be the digraph obtained by adding the edges $(a,z)$ and $(b,z)$ with multiplicity $1$ to the digraph $G[V(G) \setminus \{v\}]$. Then, we have
    \begin{multline}
        C(G') = C(G[V(G) \setminus \{v\}) + C(G',(a,z)) + C(G',(b,z)) = C(G) - C(G,v) + C(G, \{(a,v),(v,z)\})/2\\
        + C(G, \{(b,v),(v,z)\})/2 \ge C(G) - C(G,v) + C(G,v)/3 \ge C(G) - 4G(G)/9 = 5C(G)/9.
    \end{multline}
    Now, since $C(G') \le F_3(n-1)$, we obtain $C(G) \le 9F_3(n-1)/5$.

    \vspace{1pt}
    \noindent\textit{Case 3.1.2.} We now consider the case when $(a,z) \in E(G)$ with multiplicity $2$.
    Let $G'$ be the digraph obtained by adding the edges $(b,z)$ and $(a,y)$ with multiplicity $1$ to the digraph $G[V(G) \setminus \{v\}]$. Then, we have
    \begin{multline}
    \label{eqn:obj1}
        C(G') = C(G[V(G) \setminus \{v\}) + C(G',\{(b,z),(a,y)\}) + C(G' \setminus (a,y),(b,z)) + C(G' \setminus (b,z),(a,y))\\
        \ge C(G) - C(G,v) + x_{b,z}C(G)/2 + x_{a,y}C(G) = C(G)(1 - x_{a,z}- x_{b,y} - x_{b,z}/2).
    \end{multline}
    Further, we have
    \begin{align*}
        (x_{a,y} + x_{a,z})C(G) = C(G,(a,v)) &\le C(G,(b,v)) = (x_{b,y} + x_{b,z})C(G),\\
        2(x_{a,y} + x_{b,y})C(G) = 2C(G,(v,y)) &\le C(G,(v,z)) = (x_{a,z} + x_{b,z})C(G),\\
        3(x_{b,y} + x_{b,z}) = 3C(G,(b,v)) &\le C(G),\\
        (x_{a,z} + x_{b,z})C(G) + 2(x_{a,y} + x_{a,z})C(G) &\le C(G,(v,z)) + C(G,(a,z)) = C(G,z) \le C(G).
    \end{align*}
    Minimizing the function $\varphi(\boldsymbol{x}) = - x_{a,z} - x_{b,y} - x_{b,z}/2$ subject to the above constraints along with the constraint $\boldsymbol{x} \in [0,1]^4$, we obtain $\varphi(\boldsymbol{x}) \ge -9/16$, which when used in \eqref{eqn:obj1} yields $C(G') \ge 7C(G)/16$. Therefore, $C(G) \le 16F_3(n-1)/7$.

    \vspace{1pt}
    \noindent\textit{Case 3.1.3.} We now consider the case when $(b,z) \in E(G)$ with multiplicity $2$.
    Let $G'$ be the digraph obtained by adding the edges $(a,z)$ and $(b,y)$ with multiplicity $1$ to the digraph $G[V(G) \setminus \{v\}]$. Then, we have
    \begin{multline}
    \label{eqn:obj2}
        C(G') = C(G[V(G) \setminus \{v\}) + C(G',\{(a,z),(b,y)\}) + C(G' \setminus (b,y),(a,z)) + C(G' \setminus (a,z),(b,y))\\
        \ge C(G) - C(G,v) + x_{a,z}C(G)/2 + x_{b,y}C(G) = C(G)(1 - x_{a,y} - x_{a,z}/2 - x_{b,z}).
    \end{multline}
    Similar to the previous case, we have
    \begin{align*}
        (x_{a,y} + x_{a,z})C(G) = C(G,(a,v)) &\le C(G,(b,v)) = (x_{b,y} + x_{b,z})C(G),\\
        2(x_{a,y} + x_{b,y})C(G) = 2C(G,(v,y)) &\le C(G,(v,z)) = (x_{a,z} + x_{b,z})C(G),\\
        3(x_{b,y} + x_{b,z}) = 3C(G,(b,v)) &\le C(G),\\
        (x_{a,z} + x_{b,z})C(G) + 2(x_{b,y} + x_{b,z})C(G) &\le C(G,(v,z)) + C(G,(b,z)) = C(G,z) \le C(G).
    \end{align*}
    Minimizing the function $\varphi(\boldsymbol{x}) = - x_{a,y} - x_{a,z}/2 - x_{b,z}$ subject to the above constraints along with the constraint $\boldsymbol{x} \in [0,1]^4$, we obtain $\varphi(\boldsymbol{x}) \ge -11/20$, which when used in \eqref{eqn:obj2} yields $C(G') \ge 9C(G)/20$. Therefore, $C(G) \le 20F_3(n-1)/9$.
    
    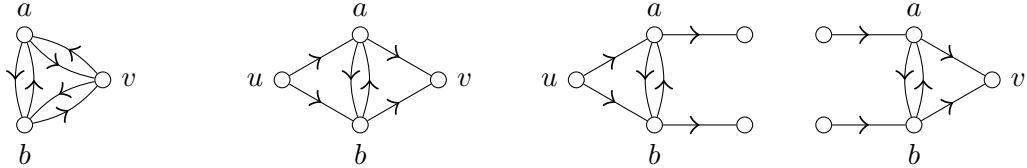
\begin{figure}[!b]
    \centering
    \begin{subfigure}{.22\textwidth}
    \[\begin{tikzpicture}[x=0.6cm, y=0.6cm,
        every edge/.style={
            draw,
            postaction={decorate,
                        decoration={markings,mark=at position 0.5 with {\arrow[scale=2]{>}}}
                       }
            }
    ]
    
    \vertex (v1) at  (0,1) [label=above:$a$]{};
    \vertex (v2) at  (0,-1) [label=below:$b$]{};
    \vertex (v3) at  (1.732,0) [label=right:$v$]{};
    
    \path
        (v1) edge[bend right=18] (v2)
        (v2) edge[bend right=18] (v1)
        
        (v1) edge[bend right=18] (v3)
        (v3) edge[bend right=18] (v1)
    
        (v2) edge[bend right=18] (v3)
        (v3) edge[bend right=18] (v2)
    ;
    \end{tikzpicture}\]
    \end{subfigure}
    \begin{subfigure}{.22\textwidth}
    \[\begin{tikzpicture}[x=0.6cm, y=0.6cm,
        every edge/.style={
            draw,
            postaction={decorate,
                        decoration={markings,mark=at position 0.5 with {\arrow[scale=2]{>}}}
                       }
            }
    ]
    
    \vertex (v1) at  (0,1) [label=above:$a$]{};
    \vertex (v2) at  (0,-1) [label=below:$b$]{};
    \vertex (v3) at  (1.732,0) [label=right:$v$]{};
    \vertex (v4) at (-1.732,0)
    [label=left:$u$]{};
    
    \path
        (v1) edge[bend right=18] (v2)
        (v2) edge[bend right=18] (v1)
        
        (v1) edge (v3)
        (v2) edge (v3)
        
        (v4) edge (v1)
        (v4) edge (v2)
    ;
    \end{tikzpicture}\]
    \end{subfigure}
    \begin{subfigure}{.22\textwidth}
    \[\begin{tikzpicture}[x=0.6cm, y=0.6cm,
        every edge/.style={
            draw,
            postaction={decorate,
                        decoration={markings,mark=at position 0.5 with {\arrow[scale=2]{>}}}
                       }
            }
    ]
    
    \vertex (v1) at  (0,1) [label=above:$a$]{};
    \vertex (v2) at  (0,-1) [label=below:$b$]{};
    \vertex (v3) at  (2,1) []{};
    \vertex (v5) at  (2,-1) []{};
    \vertex (v4) at (-1.732,0)
    [label=left:$u$]{};
    
    \path
        (v1) edge[bend right=18] (v2)
        (v2) edge[bend right=18] (v1)
        
        (v1) edge (v3)
        (v2) edge (v5)
        
        (v4) edge (v1)
        (v4) edge (v2)
    ;
    \end{tikzpicture}\]
    \end{subfigure}
    \begin{subfigure}{.22\textwidth}
    \[\begin{tikzpicture}[x=0.6cm, y=0.6cm,
        every edge/.style={
            draw,
            postaction={decorate,
                        decoration={markings,mark=at position 0.5 with {\arrow[scale=2]{>}}}
                       }
            }
    ]
    
    \vertex (v1) at  (0,1) [label=above:$a$]{};
    \vertex (v2) at  (0,-1) [label=below:$b$]{};
    \vertex (v3) at  (1.732,0) [label=right:$v$]{};
    \vertex (v4) at (-2,1)
    []{};
    \vertex (v5) at (-2,-1)
    []{};
    
    \path
        (v1) edge[bend right=18] (v2)
        (v2) edge[bend right=18] (v1)
        
        (v1) edge (v3)
        (v2) edge (v3)
        
        (v4) edge (v1)
        (v5) edge (v2)
    ;
    \end{tikzpicture}\]
    \end{subfigure}
    \caption{The cases 3.2.1, 3.2.2, 3.2.3, and 3.2.4 (from left to right).}
    \label{fig:case-3.2-subcases}
\end{figure}
    
    We shall henceforth assume that each vertex of $G$ has one incoming edge with multiplicity $k-1$ and the other with multiplicity $1$.

    \vspace{2pt}
    \noindent\textit{Case 3.2.} We consider the case in which $\text{Skel}(G)$ contains a subgraph isomorphic to $H_1$ and/or $H_2$. As shown in Figure~\ref{fig:case-3.2-subcases}, we consider the following mutually disjoint exhaustive set of cases:

    \vspace{1pt}
    \noindent\textit{Case 3.2.1.} $G$ contains distinct vertices $v, a$, and $b$ such that $(v,a), (v,b), (a,b), (b,a), (a,v), (b,v) \in E(G)$. In this case, we have $C(G) = C(G[V(G)-\{a,b,v\}]) + (k-1)^3 + 3(k-1) + 1 \le F_k(n-3) + (k-1)^3 + 3(k-1) + 1$.

    \vspace{1pt}
    \noindent\textit{Case 3.2.2.} $G$ contains distinct vertices $u,v, a, b$ such that $(u,a), (u,b), (a,b), (b,a), (a,v), (b,v) \in E(G)$. In this case, we have $C(G) \le ((k-1)^3 + 2(k-1) + 1) C(G') + (k-1) \le ((k-1)^3 + 2(k-1) + 1) F_k(n-3) + (k-1)$, where $G'$ is the digraph obtained by merging the vertices $u,v,a$ and $b$ in $G$.

    \vspace{1pt}
    \noindent\textit{Case 3.2.4.} $G$ contains distinct vertices $v, a$, and $b$ such that $(a,v), (b,v), (a,b), (b,a) \in E(G)$ but there does not exist any vertex $u$ such that $(u,a), (u,b) \in E(G)$. In this case, we get $C(G) \le ((k-1)^2 + 1)C(G') + (k-1) \le ((k-1)^2 + 1)F_k(n-2) + (k-1)$, where $G'$ is the digraph obtained by merging the vertices $a, b$ and $v$ in $G$.

    \vspace{2pt}
    \noindent\textit{Case 3.3.} We now consider the case when $\text{Skel}(G)$ is $H_1$-free and $H_2$-free. For any vertex $v \in V(\text{Skel}(G))$, let us define $E_v = \{(x,y): (x,v) \in E(\text{Skel}(G)), (x,y) \in E(\text{Skel}(G)), v \ne y\}$. Note that $|E_v| = 2$ and $E_v \cap E_u = \phi$ for any two distinct vertices $u,v \in V(\text{Skel}(G))$. Therefore, the sets $E_v$ form a uniform $n$-partition of $E(\text{Skel}(G))$.

    \vspace{1pt}
    \noindent\textit{Case 3.3.1.} If $\text{Skel}(G)$ contains exactly $n$ edges that are not in-contractible, then by Lemma~\ref{lem:structure}, each connected component of $\text{Skel}(G)$ must be isomorphic to $G_m$ for some $m \ge 4$. Now, it is easy to check that $G_{m,k}$ and $G'_{m,k}$ are the only digraphs whose skeleton is isomorphic to $G_m$ with each vertex in the digraph incident to two incoming (outgoing) edges, one with multiplicity $1$ and the other with multiplicity $k-1$. Therefore, each connected component of $G$ is isomorphic to one of $G_{m,k}$ and $G'_{m,k}$ for some $m \ge 4$. Finally, from Lemma~\ref{lem:bound}, we have $C(G) \le (k-1)F_k(n-1) + F_k(n-2)$.

    \vspace{1pt}
    \noindent\textit{Case 3.3.2.} If $\text{Skel}(G)$ contains strictly less than $n$ edges that are not in-contractible, then we can find a vertex $x \in V(\text{Skel}(G))$ such that both edges in $E_x$ are in-contractible. Let $(v,x)$ and $(w,x)$ be the incoming edges to $x$ in $G$. Let us assume that $(v,y)$ and $(w,z)$ are the other outgoing edges from $v$ and $w$, respectively. Without loss of generality, we may assume that $(v,x)$ has multiplicity $1$ in $G$. The number of cycles not passing through $(v,x)$ in $G$ is less than or equal to $(k-1)C(G')$, where $G'$ is the digraph obtained from $G$ by deleting the edge $(v,x)$ followed by contracting the edge $(v,y)$. The number of cycles passing through $(v,x)$ in $G$ is equal to the number of cycles passing through $v$ in the digraph $G''$ formed by deleting the edges $(v,y)$ and $(w,x)$ from $G$, which is further less than or equal to $C(G''')$, where $G'''$ is the digraph obtained by contracting the edges $(w,z)$ and $(v,x)$ in $G'$. Thus, we obtain $C(G) \le (k-1)C(G') + C(G''') \le (k-1)F_k(n-1) + F_k(n-2)$.

    \vspace{3pt}
    Combining all the recurrent upper bounds obtained through the proof, we obtain
    \begin{multline}
        F_k(n) \le \max(\{(k-1)F_k(n-1) + F_k(n-2), ((k-1)^2+1)F_k(n-2) + (k-1),\\ ((k-1)^3 + 2(k-1) + 1)F_k(n-3) + (k-1), F_k(n-3) + (k-1)^3 + 3(k-1) + 1\})
    \end{multline}
    for $k \ge 2, k \ne 3$. Now, we will show that $F_k(n) \le 5 \alpha(k)^n$, where $\alpha(k)$ is as defined in \eqref{eqn:alpha-k}. Using the expressions for $F_k(0), F_k(1)$ and $F_k(2)$ from the beginning of this proof, it is easy to check that $F_k(n) \le 5 \alpha(k)^n$ for $n = 0,1,2$. Now, assuming that the result holds for all natural numbers up to $n-1$, we have
    \begin{equation*}
        (k-1)F_k(n-1) + F_k(n-2) \le 5(k-1) \alpha(k)^{n-1} + 5 \alpha(k)^{n-2} = 5 \alpha(k)^n.
    \end{equation*}
    Further, it is easy to check that if the inequality
    \begin{equation*}
        ((k-1)^2+1)5 \alpha(k)^{n-2} + (k-1) \le 5 \alpha(k)^{n}
    \end{equation*}
    holds for $n = 3$, then it holds for all $n \ge 3$ since $\alpha(k) \ge 1$. And it is also easy to verify that the result indeed holds for $n = 3$. Therefore, we have
    \begin{equation*}
        ((k-1)^2+1)F_k(n-2) + (k-1) \le ((k-1)^2+1) 5 \alpha(k)^{n-2} + (k-1) \le 5 \alpha(k)^{n}
    \end{equation*}
    for all $n \ge 3$. Similar results can be shown for $F_k(n) \le ((k-1)^3 + 2(k-1) + 1)F_k(n-3) + (k-1)$ and $F_k(n) \le F_k(n-3) + (k-1)^3 + 3(k-1) + 1$. Hence, we obtain the result $F_k(n) \le 5 \alpha(k)^n$ for all non-negative integers $n$ and $k$ with $k \ge 2, k \ne 3$. Recall that for $k = 3$, we have the extra inequality $F_3(n) \le 16F_3(n-1)/7$. But since $16 \alpha(3)^{n-1}/7 \le \alpha(3)^n$, the result holds for $k = 3$ as well. Note that the constant $5$ in this result can be improved; however, we do not concern ourselves with finding the best possible constant. Finally, since $F_k(n) \ge C(G_{n,k}) \ge \alpha(k)^n$, we obtain $F_k(n) = \Theta(\alpha(k)^n)$. The constant $\alpha(2) = (1+\sqrt{5})/2$ is the golden ratio and hence $F_2(n) = \Theta(F_n)$, where $F_n$ is the $n$-th Fibonacci number.
\endproof

\begin{remark}
    It is possible to show that $F_k(n) \le (k-2+6^{1/3})^n$ using the star bound (Soules~\cite{Soules}), a generalization of Br\`egman's theorem. However, this bound is exponentially weaker than the one we have proved in Theorem~\ref{thm:cycle-bound}, which is tight up to a known constant.
\end{remark}

%% file: dmdp.tex
\section{Bounds for policy iteration on deterministic MDPs}
\label{sec:DMDPs}

We begin with the formal definition of a DMDP and then discuss the structure of the policy space for DMDPs.

\begin{definition}
    A DMDP is an MDP $(S, A, T, R, \gamma)$ with a deterministic transition function: that is, the codomain of $T$ is $\{0,1\}$.
\end{definition}

\begin{figure}[!b]
\centering
\begin{subfigure}{.32\textwidth}
\[\begin{tikzpicture}[x=0.9cm, y=0.9cm,
    every edge/.style={
        draw,
        postaction={decorate,
                    decoration={markings,mark=at position 0.5 with {\arrow[scale=2]{>}}}
                   }
        },
    every loop/.style={}
]

\vertex (v1) at  (0.62,-1.9) [label=below:$s_1$]{};
\vertex (v2) at  (-0.62,-1.9) [label=below:$s_2$]{};
\vertex (v3) at  (-1.62,-1.18) [label=left:$s_3$]{};
\vertex (v4) at  (-2,0) [label=left:$s_4$]{};
\vertex (v5) at  (-1.62,1.18) [label=left:$s_5$]{};
\vertex (v6) at  (-0.62,1.9) [label=above:$s_6$]{};
\vertex (v7) at  (0.62,1.9) [label=above:$s_7$]{};
\vertex (v8) at  (1.62,1.18) [label=right:$s_8$]{};
\vertex (v9) at  (2,0) [label=right:$s_9$]{};
\vertex (v10) at  (1.62,-1.18) [label=right:$s_{10}$]{};

\path
    (v1) edge[bend left=18] (v2)
    (v2) edge[dashed, color=red, bend left=18] (v3)
    (v3) edge[bend left=18] (v4)
    (v4) edge[dashed, color=red, bend left=18] (v5)
    (v5) edge[bend left=18] (v6)
    (v6) edge[bend right=18] (v7)
    (v7) edge[dashed, color=red, bend left=18] (v8)
    (v8) edge[dashed, color=red, bend left=18] (v9)
    (v9) edge[bend left=18] (v10)
    (v10) edge[bend left=18] (v1)
    
    (v6) edge[dashed, bend left=18, color=red] (v4)
    (v1) edge[dashed, bend right=30, color=red] (v3)
    (v10) edge[dashed, bend left=18, color=red] (v9)
    (v7) edge[bend right=18] (v6)
    (v2) edge[in=120,out=45,loop] (v2)
    (v3) edge[dashed, bend right=18, color=red] (v4)
    (v4) edge[bend right=18] (v7)
    (v5) edge[dashed, color=red, bend right=18] (v10)
    (v8) edge[bend right=18] (v10)
    (v9) edge[dashed, color=red, bend left=18] (v7)
;
\end{tikzpicture}\]
\end{subfigure}
\begin{subfigure}{.32\textwidth}
\[\begin{tikzpicture}[x=0.9cm, y=0.9cm,
    every edge/.style={
        draw,
        postaction={decorate,
                    decoration={markings,mark=at position 0.5 with {\arrow[scale=2]{>}}}
                   }
        }
]

\vertex (v1) at  (0.62,-1.9) [label=below:$s_1$]{};
\vertex (v2) at  (-0.62,-1.9) [label=below:$s_2$]{};
\vertex (v3) at  (-1.62,-1.18) [label=left:$s_3$]{};
\vertex (v4) at  (-2,0) [label=left:$s_4$]{};
\vertex (v5) at  (-1.62,1.18) [label=left:$s_5$]{};
\vertex (v6) at  (-0.62,1.9) [label=above:$s_6$]{};
\vertex (v7) at  (0.62,1.9) [label=above:$s_7$]{};
\vertex (v8) at  (1.62,1.18) [label=right:$s_8$]{};
\vertex (v9) at  (2,0) [label=right:$s_9$]{};
\vertex (v10) at  (1.62,-1.18) [label=right:$s_{10}$]{};

\path
    (v2) edge[dashed, color=red, bend left=18] (v3)
    (v3) edge[bend left=18] (v4)
    (v4) edge[dashed, color=red, bend left=18] (v5)
    (v5) edge[bend left=18] (v6)
    (v8) edge[dashed, color=red, bend left=18] (v9)
    (v9) edge[bend left=18] (v10)
    
    (v6) edge[dashed, bend left=18, color=red] (v4)
    (v1) edge[dashed, bend right=30, color=red] (v3)
    (v10) edge[dashed, bend left=18, color=red] (v9)
    (v7) edge[bend right=18] (v6)
;
\end{tikzpicture}\]
\end{subfigure}
\begin{subfigure}{.32\textwidth}
\[\begin{tikzpicture}[x=0.9cm, y=0.9cm,
    every edge/.style={
        draw,
        postaction={decorate,
                    decoration={markings,mark=at position 0.5 with {\arrow[scale=2]{>}}}
                   }
        }
]

\vertex (v2) at  (-0.62,-1.9) [label=below:$s_2$]{};
\vertex (v3) at  (-1.62,-1.18) [label=left:$s_3$]{};
\vertex (v4) at  (-2,0) [label=left:$s_4$]{};
\vertex (v5) at  (-1.62,1.18) [label=left:$s_5$]{};
\vertex (v6) at  (-0.62,1.9) [label=above:$s_6$]{};
\vertex (v8) at  (1.62,1.18) [label=right:$s_8$]{};
\vertex (v9) at  (2,0) [label=right:$s_9$]{};
\vertex (v10) at  (1.62,-1.18) [label=right:$s_{10}$]{};

\path
    (v2) edge[dashed, color=red, bend left=18] (v3)
    (v3) edge[bend left=18] (v4)
    (v4) edge[dashed, color=red, bend left=18] (v5)
    (v5) edge[bend left=18] (v6)
    (v6) edge[dashed, bend left=18, color=red] (v4)
    (v8) edge[dashed, color=red, bend left=18] (v9)
    (v9) edge[bend left=18] (v10)
    (v10) edge[dashed, bend left=18, color=red] (v9)
;
\end{tikzpicture}\]
\end{subfigure}
\caption{An example of a directed multigraph corresponding to a $10$-state $2$-action DMDP, the policy digraph $G_\pi$ for $\pi = 0010101010$, and the path-cycles $P^\pi_{s_2}$ and $P^\pi_{s_8}$ (from left to right), where the dashed (red) edges and solid (black) edges correspond to actions $0$ and $1$ respectively.}
\label{fig:dmdp-digraph}
\end{figure}
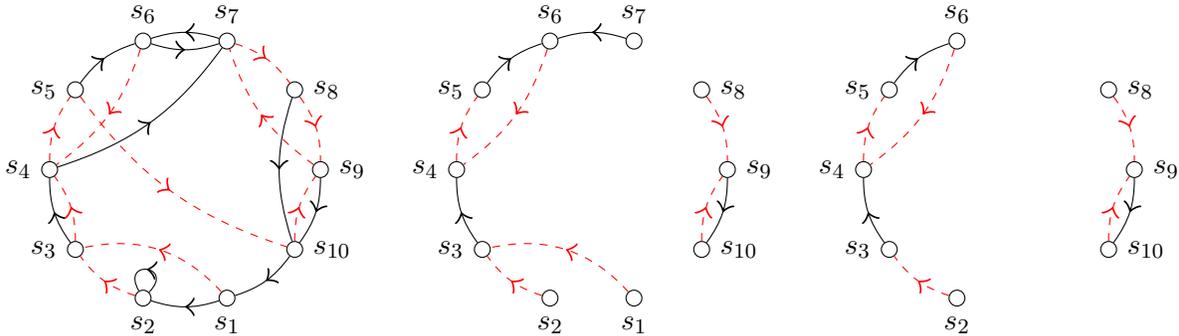

A DMDP $M$ can be viewed as a directed multigraph $G_M$ with $S$ as the set of vertices, which contains an edge corresponding to each state-action pair. Further, a policy $\pi \in \Pi_M$ can be viewed as a digraph $G_\pi$ in which each vertex has outdegree one. Such digraphs are known as functional graphs since they correspond to functions defined on the set of vertices. A functional graph is a union of its connected components, each containing a single cycle and paths leading into the cycle. See Figure~\ref{fig:dmdp-digraph} for an example. Post and Ye's~\cite{PostYe} analysis suggests that the value function of any policy is primarily dictated by the cycles present in the corresponding digraph. We now describe an alternative, slightly different way to view policies for DMDPs.

\begin{definition}
    A digraph isomorphic to $G = (\{v_1, \dots, v_n\}, \{(v_i, v_{i+1}): 1 \le i < n\} \cup \{(v_n, v_m)\})$ for some $m,n \in \mathbb{N}$ with $m \le n$ is called a path-cycle.
\end{definition}

Let $M$ be a DMDP with its set of states $S = \{s_1, \dots, s_n\}$. A policy $\pi \in \Pi_M$ can be viewed as an $n$-tuple $(P^\pi_{s_1}, \dots, P^\pi_{s_n})$ of path-cycles, where $P^\pi_{s}$ is the path-cycle obtained by following $\pi$ starting from state $s$. We shall call $(P^\pi_{s_1}, \dots, P^\pi_{s_n})$ the (path-cycle) representation of policy $\pi$. Note that $V^{\pi}(s)$ is completely determined by the corresponding path-cycle $P^\pi_s$ for any $s \in S$.

\begin{lemma}
\label{lem:new-path-cycles}
    Let $M$ be a DMDP and $\pi_1 \prec \dots \prec \pi_\ell$ be an increasing sequence of policies in $\Pi_M$. Then, for each $1 < i \le \ell$, the representation of $\pi_i$ contains a path-cycle which is not part of representations of $\pi_j$ for any $1 \le j < i$. Therefore, we can associate a possibly non-unique sequence of distinct path-cycles to any increasing sequence of policies.
\end{lemma}
\proof
    Let $1 < i \le \ell$. Since $\pi_{i-1} \prec \pi_i$, we have $V^{\pi_{i-1}}(s) < V^{\pi_i}(s)$ for some state $s \in S$. Now since $\pi_j \preceq \pi_{i-1}$ for $1 \le j < i$, $V^{\pi_j}(s) < V^{\pi_i}(s)$ for each $1 \le j < i$. If $P^{\pi_i}_s$ were a part of the path-cycle representation of $\pi_j$ for some $1 \le j < i$, then $V^{\pi_j}(s) = V^{\pi_i}(s)$, a contradiction.
\endproof

Before we prove results about the number of steps PI takes to converge, we establish some graph theoretic notation and results, which shall be useful later.

For any digraph $G$, we denote the number of path-cycles and paths in $G$ by $C'(G)$ and $C''(G)$, respectively. Clearly, $C(G) \le C'(G)$ since every cycle is a path-cycle.

For integers $n \ge 0$, $k \ge 2$, we define $\mathcal{G}^{n,k}$ as the set of all digraphs with $n$ vertices and outdegree $k$ (we allow digraphs to contain loops and multi-edges, as mentioned in Section~\ref{sec:cycles}). Note that if $M$ is an $n$-state $k$-action DMDP, then $G_M \in \mathcal{G}^{n,k}$.

\begin{definition}
    For $G \in \mathcal{G}^{n,k}$, we define $N_1(G)$ to be the number of path-cycles in the digraph $G'$ obtained from $G$ by replacing each multi-edge of multiplicity $k$ with a corresponding edge of multiplicity $1$: that is, $N_1(G) = C'(G')$. Similarly, we define $N_2(G)$ to be the number of path-cycles in the skeleton of $G$: that is, $N_2(G) = C'(\text{Skel}(G))$.
\end{definition}

We prove below some bounds on $N_1(G)$ and $N_2(G)$ using the bounds on the number of cycles established in Section~\ref{sec:cycles}.

\begin{lemma}
\label{lem:num-path-cycles-nk}
    Let $G \in \mathcal{G}_{\text{simple}}^{n, k}$. Then $C'(G) \le n^2 k (k+1)!^{(n-1)/(k+1)}$.
\end{lemma}
\proof
    A path-cycle in $G$ can be viewed as a pair consisting of a path and an edge from the terminal vertex of the path to a vertex in the path. Therefore, $C'(G) \le k C''(G)$. Let $s,t \in V(G)$ and the number of paths from $s$ to $t$ in $G$ be denoted by $C''_{s,t}(G)$. Then $C''_{s,t}(G) \le C(G')$, where $G'$ is the digraph obtained by deleting all incoming edges to $s$ and outgoing edges from $t$ in $G$ followed by merging vertices $s$ and $t$. It is easy to check that $G' \in \mathcal{G}_{\text{simple}}^{n-1,k}$. Therefore, we have $C(G') \le (k+1)!^{(n-1)/(k+1)}$ from Theorem~\ref{thm:nk}. Hence, $C''_{s,t}(G) \le (k+1)!^{(n-1)/(k+1)}$. Summing over all pairs of vertices $s,t \in V(G)$, we obtain $C''(G) \le n^2 (k+1)!^{(n-1)/(k+1)}$, which combined with $C'(G) \le k C''(G)$ yields the desired result.
\endproof

\begin{lemma}
\label{lem:num-path-cycles}
    Let $G \in \mathcal{G}_{\text{multi}}^{n, k}$. Then $C'(G) \le 5 n^2 k \alpha(k)^{n-1}$.
\end{lemma}
\proof
    A path-cycle in $G$ can be viewed as a pair consisting of a path and an edge from the terminal vertex of the path to a vertex in the path. Therefore, $C'(G) \le k C''(G)$. Let $s,t \in V(G)$ and the number of paths from $s$ to $t$ in $G$ be denoted by $C''_{s,t}(G)$. Then $C''_{s,t}(G) \le C(G')$, where $G'$ is the digraph obtained by deleting all incoming edges to $s$ and outgoing edges from $t$ in $G$ followed by merging vertices $s$ and $t$. Note that some vertices in $G'$ could have outdegree less than $k$. We add a sufficient number of self-loops to each such vertex so that the outdegree of each vertex is equal to $k$ in the resulting digraph $G''$, which satisfies $C(G') \le C(G'')$. Since $G'' \in \mathcal{G}_{\text{multi}}^{n, k}$, we have $C(G'') \le 5 \alpha(k)^{n-1}$ from Theorem~\ref{thm:cycle-bound}. Therefore, $C''_{s,t}(G) \le 5 \alpha(k)^{n-1}$. Summing over all pairs of vertices $s,t \in V(G)$, we obtain $C''(G) \le 5 n^2 \alpha(k)^{n-1}$, which combined with $C'(G) \le k C''(G)$ yields the desired result.
\endproof

\begin{proposition}
\label{prop:bound}
    Let $G \in \mathcal{G}^{n,k}$. Then we have
    \begin{enumerate}[(1)]
        \item $N_1(G) \le 5n^2 k \alpha(k)^{n-1}$, and
        \item $N_2(G) \le n^2 k (k+1)!^{(n-1)/(k+1)}$.
    \end{enumerate}
\end{proposition}

\proof
    Suppose that $G \in \mathcal{G}^{n,k}$. Let $G'$ be the digraph obtained from $G$ by replacing each multi-edge of multiplicity $k$ with a corresponding edge of multiplicity $1$. Note that some vertices in the digraph $G'$ could have outdegree less than $k$. We add a sufficient number of self-loops to each such vertex so that the outdegree of each vertex is equal to $k$ in the resulting digraph $G''$, which satisfies $C'(G') \le C'(G'')$. Now, from Lemma~\ref{lem:num-path-cycles}, we obtain $C'(G'') \le 5n^2 k \alpha(k)^{n-1}$ since $G'' \in \mathcal{G}^{n,k}_{multi}$. Therefore, $N_1(G) = C'(G') \le C'(G'') \le 5n^2 k \alpha(k)^{n-1}$. This finishes the proof of (1).
    
    Now suppose that $G \in \mathcal{G}^{n,k}$. Note that Skel$(G) \in \mathcal{G}^{n,k}_{simple}$. Therefore, by Lemma~\ref{lem:num-path-cycles-nk}, we obtain $N_2(G) = C'(\text{Skel}(G)) \le n^2 k (k+1)!^{(n-1)/(k+1)}$. This finishes the proof of (2).
\endproof

\subsection{Policy iteration with arbitrary action selection.} We begin by defining an equivalence relation on the edges of a DMDP digraph and an induced equivalence relation on the path-cycles therein.

\begin{definition}
\label{def:equiv}
    Let $M = (S, A, T, R, \gamma)$ be a DMDP. We say that a state $s \in S$ is \emph{non-branching} if there exists $s' \in S$ such that $T(s, a, s') = 1$ for all $a \in A$. For any $a_1, a_2 \in A$, we say that the edges corresponding to $(s, a_1)$ and $(s, a_2)$ in $G_M$ are equivalent if $s$ is a non-branching state. Given two path-cycles $P_1$ and $P_2$ in $G_M$, we say that $P_1 \sim P_2$ if $P_1$ and $P_2$ differ only in equivalent edges.
\end{definition}

Given a DMDP $M$, $\sim$ is an equivalence relation on the set of path-cycles in $G_M$.

Our approach for proving a bound on the number of policies visited by a PI algorithm with arbitrary action selection is as follows:
\begin{enumerate}
    \item associate a sequence of distinct path-cycles to the sequence of policies visited (by Lemma~\ref{lem:new-path-cycles}),
    \item show that this sequence does not contain many equivalent path-cycles (see Lemma~\ref{lem:equiv-path-cycles}), and
    \item use Proposition~\ref{prop:bound} to bound the number of path-cycles in the digraph obtained by identifying edges under the equivalence relation defined in Definition~\ref{def:equiv}.
\end{enumerate}

In the following lemma, we show that a sequence of distinct path-cycles associated with an increasing sequence of policies obtained via policy improvement with arbitrary action selection does not contain many equivalent path-cycles.

\begin{lemma}
\label{lem:equiv-path-cycles}
    Let $M = (S, A, T, R, \gamma)$ be an $n$-state, $k$-action DMDP and $\pi_1 \prec \dots \prec \pi_\ell$ be an increasing sequence of policies obtained via policy improvement with arbitrary action selection. Further, let $P = P^{\pi_1}_{s_1}, \dots, P^{\pi_\ell}_{s_\ell}$ be an associated sequence of path-cycles. Then, the size of each equivalence class of $P$ under the equivalence relation $\sim$ is at most $kn$.
\end{lemma}
\proof
    Let $S' \subseteq S$ be the set of non-branching states. Let $Q = P^{\pi_{i_1}}_{s}, \dots, P^{\pi_{i_m}}_{s}$ be a subsequence of $P$ consisting of equivalent path-cycles. And let $1 \le j < m$. Then $P^{\pi_{i_j}}_{s}$ and $P^{\pi_{i_{j+1}}}_{s}$ differ in edges coming out of a non-branching state, say $\pi_{i_j}(s') = a$ and $\pi_{i_{j+1}}(s') = a'$ for some $s' \in S'$ and distinct actions $a, a' \in A$. We remark that $P^{\pi_{i_j}}_{s}$ and $P^{\pi_{i_{j+1}}}_{s}$ may additionally differ in other equivalent edges. Let $P^{\pi}_{s}$ be a path-cycle that appears after $P^{\pi_{i_j}}_{s}$ in $Q$. We claim that $\pi(s') \ne a$.
    
    To see this, note that there exists some $i' \in [i_j, i_{j+1})$ such that the action at state $s'$ is switched from $a$ to $a''$ ($\ne a$) while going from $\pi_{i'}$ to $\pi_{i'+1}$. Therefore, $\rho^{\pi_{i'}}(s',a'') > \rho^{\pi_{i'}}(s',a)$. But since $s' \in S'$, this implies $R(s',a'') > R(s',a)$. Since $s'$ is a non-branching state, this further implies that action $a$ is not an improving action for any policy that appears after $\pi_{i'}$ in the sequence $\pi_1, \dots, \pi_\ell$. Using an inductive argument, it follows that action $a$ is not taken at the state $s'$ in any policy after $\pi_{i'}$. In particular, this implies $\pi(s') \ne a$.
    
    Thus, each time we transition along $Q$ from one path-cycle to the next, one edge is eliminated from $G_M$ in the sense that it cannot be a part of subsequent path-cycles in $Q$. Hence, the number of path-cycles in $Q$ is bounded above by the number of edges in $G_M$, which equals $kn$.
\endproof

In Theorem~\ref{thm:general-length-longest-path}, we show that the length of any increasing sequence of policies obtained via policy improvement with arbitrary action selection (as in Lemma~\ref{lem:equiv-path-cycles}) for DMDP $M$ is bounded above by $N_1(G_M)$ up to a multiplicative factor which is polynomial in $n$ and $k$. Equivalently, one may view the following theorem as providing an upper bound on the length of the longest directed path in the PI-DAG of the DMDP $M$.

\begin{theorem}
\label{thm:general-length-longest-path}
    Let $M$ be an $n$-state, $k$-action DMDP and let $\pi_1 \prec \pi_2 \prec \dots \prec \pi_\ell$ be an increasing sequence of policies obtained via policy improvement with arbitrary action selection on $M$. Then $\ell \le kn N_1(G_M)$.
\end{theorem}
\proof
    Let $\pi_1 \prec \pi_2 \prec \dots \prec \pi_\ell$ be an increasing sequence of policies obtained via policy improvement with arbitrary action selection on $M$. And let $P = P^{\pi_1}_{s_1}, \dots, P^{\pi_\ell}_{s_\ell}$ be an associated sequence of path-cycles. Then, the number of path-cycles in $P$ equals the sum of the sizes of the equivalence classes of $P$ under $\sim$. Using Lemma~\ref{lem:equiv-path-cycles}, we obtain
    \begin{align*}
        \ell &\le kn \times \text{the number of equivalence classes of $P$ under $\sim$},\\
        &\le kn C'(G_M'),
    \end{align*}
    where $G_M'$ is the digraph obtained from $G_M$ by replacing each multi-edge of multiplicity $k$ with a corresponding edge of multiplicity $1$. Using $N_1(G_M) = C'(G_M')$ in the above inequality yields the desired result.
\endproof

\begin{corollary}
\label{general-PI-upper}
    Any run of a PI algorithm with arbitrary action selection on an $n$-state, $k$-action DMDP terminates in at most $5 n^3 k^2 \alpha(k)^{n-1} = O(n^3 k \alpha(k)^n)$ steps.
\end{corollary}
\proof
    Let $M$ be an $n$-state, $k$-action DMDP and $\pi_1 \prec \pi_2 \prec \dots \prec \pi_\ell$ be the sequence of policies in $\Pi_M$ encountered during a run of a PI algorithm with arbitrary action selection on $M$. Then, using Theorem~\ref{thm:general-length-longest-path}, we obtain $\ell \le kn N_1(G_M)$. Further, we have $N_1(G_M) \le 5n^2 k \alpha(k)^{n-1}$ from Proposition~\ref{prop:bound} (1), which implies $\ell \le 5n^3 k^2 \alpha(k)^{n-1}$. Finally, since $\alpha(k) \ge k/2$, we obtain $\ell = O(n^3 k \alpha(k)^n)$.
\endproof

\subsection{Policy iteration with max-gain action selection.} We begin by defining an equivalence relation on the edges of a DMDP digraph and an induced equivalence relation on the path-cycles therein.

\begin{definition}
\label{def:equiv2}
    Let $M = (S, A, T, R, \gamma)$ be a DMDP. For $s \in S$ and $a_1, a_2 \in A$, we say that the edges corresponding to $(s, a_1)$ and $(s, a_2)$ in $G_M$ are equivalent if there exists $s' \in S$ such that $T(s,a_1,s') = T(s,a_2,s') = 1$. Given two path-cycles $P_1$ and $P_2$ in $G_M$, we say that $P_1 \approx P_2$ if $P_1$ and $P_2$ differ only in equivalent edges.
\end{definition}

Given a DMDP $M$, $\approx$ is an equivalence relation on the set of path-cycles in $G_M$. We define a stronger notion of equivalence between edges below, which shall be useful in the proof of Lemma~\ref{lem:equiv-path-cycles-max}.

\begin{definition}
    For $s \in S$ and $a_1, a_2 \in A$, we say that the edges corresponding to $(s,a_1)$ and $(s,a_2)$ are quasi-equal, denoted $(s,a_1) \equiv (s,a_2)$, if $(s,a_1)$ and $(s,a_2)$ are equivalent in the sense of Definition~\ref{def:equiv2} and $R(s,a_1) = R(s,a_2)$.
\end{definition}

Let $E_{\text{max}}$ be the set of state-action pairs with the highest reward among the state-action pairs in their respective equivalence classes. Note that if $(s,a) \in E_{\text{max}}$, then the intersection of $E_{\text{max}}$ with the equivalence class of $(s,a)$ is equal to the quasi-equality class of $(s,a)$.

We use the same proof strategy as for PI with arbitrary action selection. In the following lemma, we show that a sequence of distinct path-cycles associated with an increasing sequence of policies obtained via policy improvement with max-gain action selection does not contain many equivalent path-cycles.

\begin{lemma}
\label{lem:equiv-path-cycles-max}
    Let $M = (S, A, T, R, \gamma)$ be an $n$-state, $k$-action DMDP and $\pi_1 \prec \pi_2 \prec \dots \prec \pi_\ell$ be an increasing sequence of policies obtained via policy improvement with max-gain action selection on $M$. Further, let $P = P^{\pi_1}_{s_1}, \dots, P^{\pi_\ell}_{s_\ell}$ be an associated sequence of path-cycles. Then, the size of each equivalence class of $P$ under the equivalence relation $\approx$ is at most $n+1$.
\end{lemma}

\proof
    Let $Q = P^{\pi_{i_1}}_{s}, \dots, P^{\pi_{i_m}}_{s}$ be a subsequence of $P$ consisting of equivalent path-cycles. Let $1 \le j < m$. Then $P^{\pi_{i_j}}_{s}$ and $P^{\pi_{i_{j+1}}}_{s}$ differ in equivalent but not quasi-equal edges (due to strict improvement in the value function) coming out of a state, say $(s',a_1)$ and $(s',a_2)$, respectively, for some $s' \in S$ and distinct actions $a_1, a_2 \in A$. We remark that $P^{\pi_{i_j}}_{s}$ and $P^{\pi_{i_{j+1}}}_{s}$ may additionally differ in other equivalent edges. Let $P^{\pi}_{s}$ be a path-cycle that appears after $P^{\pi_{i_j}}_{s}$ in $Q$. We claim that $(s',\pi(s')) \equiv (s',a_2)$.
    
    To see this, note that there exists some $i' \in [i_j, i_{j+1})$ such that the action at state $s'$ is switched to $a_2$ while going from $\pi_{i'}$ to $\pi_{i'+1}$. Since we are considering policy improvement with max-gain action selection, any edge that is switched to must be contained in $E_{\text{max}}$. Therefore, $(s',a_2) \in E_{\text{max}}$. Similarly, for any subsequent path-cycle $P^{\pi}_{s}$ in $Q$, we have $(s',\pi(s')) \in E_{\text{max}}$. Further, the edges $(s',a_2)$ and $(s',\pi(s'))$ must be equivalent since all path-cycles in $Q$ are equivalent. Therefore, we conclude that $(s',\pi(s')) \equiv (s',a_2)$.
    
    Thus, each time we transition along $Q$ from one path-cycle to the next, the action at one state becomes fixed in the sense that the edge coming out of that state cannot change its quasi-equality class in any subsequent transitions along the subsequence $Q$. Hence, there can be at most $n$ transitions, implying $m \le n+1$.
\endproof

In Theorem~\ref{thm:general-length-longest-path-max}, we show that the length of any increasing sequence of policies obtained via policy improvement with max-gain action selection (as in Lemma~\ref{lem:equiv-path-cycles-max}) for DMDP $M$ is bounded above by $N_2(G_M)$ up to a multiplicative factor which is polynomial in $n$ and $k$.

\begin{theorem}
\label{thm:general-length-longest-path-max}
    Let $M$ be an $n$-state, $k$-action DMDP and let $\pi_1 \prec \pi_2 \prec \dots \prec \pi_\ell$ be an increasing sequence of policies obtained via policy improvement with max-gain action selection on $M$. Then $\ell \le (n+1) N_2(G_M)$.
\end{theorem}
\proof
    Let $\pi_1 \prec \pi_2 \prec \dots \prec \pi_\ell$ be an increasing sequence of policies obtained via policy improvement with max-gain action selection. And let $P = P^{\pi_1}_{s_1}, \dots, P^{\pi_\ell}_{s_\ell}$ be an associated sequence of path-cycles. Then, the number of path-cycles in $P$ equals the sum of the sizes of the equivalence classes of $P$ under $\approx$. Using Lemma~\ref{lem:equiv-path-cycles-max}, we obtain
    \begin{align*}
        \ell &\le (n+1) \times \text{the number of equivalence classes of $P$ under $\approx$},\\
        &\le (n+1) C'(\text{Skel}(G_M)).
    \end{align*}
    Now using $N_2(G_M) = C'(\text{Skel}(G_M))$ in the above inequality yields the desired result.
\endproof

\begin{corollary}
\label{general-PI-upper-max}
    Any run of a PI algorithm with arbitrary action selection on an $n$-state $k$-action DMDP terminates in at most $(n+1)n^2 k (k+1)!^{(n-1)/(k+1)} = O\left( n^3 \left(\left(1+ O\left(\frac{\log k}{k}\right)\right)\frac{k}{e}\right)^n\right)$ steps.
\end{corollary}

\proof
    Let $M$ be an $n$-state, $k$-action DMDP and let $\pi_1 \prec \pi_2 \prec \dots \prec \pi_\ell$ be the sequence of policies in $\Pi_M$ encountered during a run of a PI algorithm with max-gain action selection on $M$. Then, using Theorem~\ref{thm:general-length-longest-path-max}, we obtain $\ell \le (n+1) N_2(G_M)$. Further, we have $N_2(G_M) \le n^2 k (k+1)!^{(n-1)/(k+1)}$ from Proposition~\ref{prop:bound} (2), which implies $\ell \le (n+1)n^2 k (k+1)!^{(n-1)/(k+1)}$. Now, since $n+1 \le 2n$ and $(k+1)!^{1/(k+1)} \ge (k+1)/e \ge k/e$, we obtain $\ell = O(n^3 (k+1)!^{n/(k+1)})$.
    
    We will now show that $(k+1)!^{1/(k+1)} = \left(1 + O\left(\frac{\log k}{k}\right)\right) \frac{k}{e}$ to finish the proof of this theorem.
    From Stirling's approximation, we have $k! \le e k^{k+1/2} e^{-k}$, which further yields $(k+1)! \le e (k+1) k^{k+1/2} e^{-k}$. Taking the $(k+1)$-st root on both sides, we obtain $(k+1)!^{1/(k+1)} \le \frac{k}{e} (e^2 (k+1) k^{-1/2})^{1/(k+1)}$. Now, $(e^2 (k+1) k^{-1/2})^{1/(k+1)} \le (k+1)^{1/(k+1)}$ for $k \ge e^4$. Since $(k+1)^{1/(k+1)} = 1 + O(\frac{\log k}{k})$, we conclude that $(k+1)!^{1/(k+1)} = \left(1 + O\left(\frac{\log k}{k}\right)\right) \frac{k}{e}$.
\endproof

Corollary~\ref{general-PI-upper} and Corollary~\ref{general-PI-upper-max} provide upper bounds on the number of steps required by all and max-gain policy iteration algorithms to converge for DMDPs. One of the most commonly used variants of PI is Howard's PI (with max-gain action selection). We make a few observations about this PI variant below.

\begin{remark}
    Combining the bounds in Corollary~\ref{general-PI-upper} and Corollary~\ref{general-PI-upper-max}, the tightest upper bound that we have proved for Howard's PI on $n$-state $k$-action DMDPs is of the form $O(\text{Poly}(n,k) \beta(k)^n)$, where $\beta(k)$ is given by
    \begin{equation*}
        \beta(k) = \begin{cases}
        (1+\sqrt{5})/2, &\text{if }k = 2, \\
        (k+1)!^{1/(k+1)}, &\text{for } k \ge 3.
        \end{cases}
    \end{equation*}    
\end{remark}

\begin{remark}
    The bound in the Corollary~\ref{general-PI-upper-max} applies to Howard's PI since it uses max-gain action selection. However, one can show that there are no equivalent path-cycles (except possibly the first one) in the sequence of path-cycles associated with a sequence of policies visited by Howard's PI. Consequently, the polynomial factor in the bound can be improved by a factor of $n$, and a tighter upper bound of $1 + n^2 k (k+1)!^{(n-1)/(k+1)}$ steps holds for Howard's PI.
\end{remark}

\begin{remark}
    Post and Ye~\cite{PostYe} derived several properties for the run of max-gain Simplex PI on DMDPs and used them to prove polynomial bounds on the complexity of max-gain Simplex PI. We observe that Lemmas~5, 6, and 7 in their work also hold for the run of Howard's PI on DMDPs. Therefore, at most polynomial many steps elapse between the creation of new cycles during a run of Howard's PI on a DMDP. This result, coupled with the fact that a cycle can be formed at most once during the entire run, allows one to bound above the number of steps taken by Howard's PI by the number of cycles in the skeleton of the DMDP digraph $G_M$ up to a polynomial factor. This method yields similar upper bounds for Howard's PI on DMDPs as Corollary~\ref{general-PI-upper-max}, albeit with a larger polynomial factor. However, this method cannot be used to establish bounds on the complexity of a general PI algorithm.
\end{remark}

%% file: twostate.tex
\section{Tighter bounds for MDPs with two states}
\label{sec:2statebounds}

The upper bound from Corollary~\ref{general-PI-upper}, when applied to DMDPs with $k = 2$ actions, is observed to be an exponentially vanishing fraction of the $2^{n}$ \textit{lower bound} furnished by Melekopoglou and Condon~\cite{MelekopoglouCondon} for MDPs, as $n \to \infty$. As yet, we do not have an exponentially-separated lower bound for MDPs when $k \geq 3$, whether for arbitrary or for max-gain action selection. In this section, we present a preliminary investigation of the possible separation in the running-times of PI on DMDPs and MDPs for general $k$. We analyse DMDPs and MDPs with $n = 2$ states, but with an arbitrary number of actions: that is, $k \geq 2$. We obtain the following results.

\subsection{Results.}
\label{subsec:2statebounds-results}

First, for $k \geq 2$, we show the existence of a $2$-state, $k$-action MDP whose PI-DAG contains a Hamiltonian path.

\begin{proposition}
\label{prop:2state-arbitrary-mdp-lb}
Fix $k \geq 2$. There exists a triple $(M, \pi_{0}, \mathcal{L})$, where
\begin{enumerate}
    \item $M = (S, A, T, R, \gamma)$ is an MDP with $|S| = 2$ and $|A| = k$,
    \item $\pi_{0}: S \to A$ is a policy for $M$, and 
    \item $\mathcal{L}$ is a PI algorithm
\end{enumerate}
such that if $\mathcal{L}$ is initialised at $\pi_{0}$ and executed on $M$, it visits all the $k^{2}$ policies for $M$.
\end{proposition}

On the other hand, we can upper-bound the number of vertices in paths in the PI-DAG of any $2$-state, $k$-action DMDP, $k \geq 2$, by $\frac{k^{2}}{2} + 2k - 1$.

\begin{proposition}
\label{prop:2state-arbitrary-dmdp-ub}
For every triple $(M, \pi_{0}, \mathcal{L})$ comprising
\begin{enumerate}
    \item a DMDP $M = (S, A, T, R, \gamma)$ with $|S| = 2$ and $|A| = k \geq 2$,
    \item a policy $\pi_{0}: S \to A$ for $M$, and 
    \item a PI algorithm $\mathcal{L}$,
\end{enumerate}
the number of policies evaluated by 
$\mathcal{L}$ if initialised at $\pi_{0}$ and executed on $M$ is at most $\frac{k^{2}}{2} + 2k -1$.
\end{proposition}

When we restrict our attention to PI algorithms that perform max-gain action selection, we show that on certain $2$-state, $k$-action MDPs, $k \geq 2$, a variant from the family can visit a number of policies that is linear in $k$.

\begin{proposition}
\label{prop:2state-max-gain-mdp-lb}
Fix $k \geq 2$. There exists a triple $(M, \pi_{0}, \mathcal{L})$, where
\begin{enumerate}
\item $M = (S, A, T, R, \gamma)$ is an MDP with $|S| = 2$ and $|A| = k$,
\item $\pi_{0}: S \to A$ is a policy for $M$, and 
\item $\mathcal{L}$ is a PI algorithm that performs max-gain action selection
\end{enumerate}
such that if $\mathcal{L}$ is initialised at $\pi_{0}$ and executed on $M$, it visits at least $2k - 1$ policies for $M$.
\end{proposition}

By contrast, if PI with max-gain selection is performed on a $2$-state, $k$-action DMDP, $k \geq 2$, the number of policies visited is upper-bounded by a constant.

\begin{proposition}
\label{prop:2state-max-gain-dmdp-ub}
For every triple $(M, \pi_{0}, \mathcal{L})$ comprising
\begin{enumerate}
    \item a DMDP $M = (S, A, T, R, \gamma)$ with $|S| = 2$ and $|A| = k \geq 2$,
    \item a policy $\pi_{0}: S \to A$ for $M$, and 
    \item a PI algorithm $\mathcal{L}$ that performs max-gain action selection,
\end{enumerate}
the number of policies evaluated by 
$\mathcal{L}$ if initialised at $\pi_{0}$ and executed on $M$ is at most $7$.
\end{proposition}

In summary, propositions \ref{prop:2state-arbitrary-mdp-lb} and \ref{prop:2state-arbitrary-dmdp-ub} together establish for each $k \geq 4$ the existence of $2$-state, $k$-action MDPs on which PI can visit more policies than it possibly could if transitions were constrained to be deterministic. Propositions \ref{prop:2state-max-gain-mdp-lb} and \ref{prop:2state-max-gain-dmdp-ub} affirm the same pattern when PI is restricted to max-gain action selection. In both cases, the difference between the lower bound for MDPs and the upper bound for DMDPs increases with $k$.

In the remainder of this section, we furnish proofs of propositions \ref{prop:2state-arbitrary-mdp-lb}--\ref{prop:2state-max-gain-dmdp-ub}. Separate constructions are given towards the lower bounds (propositions \ref{prop:2state-arbitrary-mdp-lb} and \ref{prop:2state-max-gain-mdp-lb}). The upper bounds (propositions \ref{prop:2state-arbitrary-dmdp-ub} and \ref{prop:2state-max-gain-dmdp-ub}) are established by noting constraints on the switches between different actions in $2$-state DMDPs.

\subsection{Proofs of lower bounds for MDPs.}
\label{subsec:2statebounds-proofsoflowerboundsformdps}

\begin{figure}[t]
\mbox{
\begin{subfigure}{.32\textwidth}
\[\begin{tikzpicture}[x=1.75cm, y=1.75cm,
    every edge/.style={
        draw,
        postaction={decorate,
                    decoration={markings,mark=at position 0.52 with {\arrow[scale=2]{>}}}
                   }
        },
    every loop/.style={},
    el/.style = {inner sep=2pt, align=left, sloped},
]

\vertex[minimum width=0.6cm,circle,inner sep=0pt] (v1) at  (0,0) {$1$};
\vertex[minimum width=0.6cm,circle,inner sep=0pt] (v2) at  (0, -1.42) {$2$};

\path (v1) edge[in=60,out=120,loop] node[el,above=1mm] {$\lambda^{1}_{a}, \mu^{1}_{a}$} (v1);

\path (v1) edge[in=75,out=-75] node[el,above,rotate=90, xshift=6ex,yshift=3ex] {$1- \lambda^{1}_{a}, \mu^{1}_{a}$} (v2);

\path (v2) edge[in=-120,out=-60,loop] node[el,below=1mm] {$\lambda^{2}_{a^{\prime}}, \mu^{2}_{a^{\prime}}$} (v2);

\path (v2) edge[in=-105,out=105] node[el,above,rotate=-90,xshift=-7ex,yshift=-6ex] {$1- \lambda^{2}_{a^{\prime}}, \mu^{2}_{a^{\prime}}$} (v1);

\end{tikzpicture}\]
\caption{Transitions in $2$-state MDP.}
\label{fig:lambda-mu}
\end{subfigure}

\begin{subfigure}{.34\textwidth}
\centering
\begin{tabular}{|c|c|c|c|c|}
\hline
24 & 23 & 19 & 13 & 5 \\ \hline    
21 & 22 & 20 & 14 & 6 \\ \hline    
16 & 17 & 18 & 15 & 7 \\ \hline    
9 & 10 & 11 & 12 & 8 \\ \hline    
0 & 1 & 2 & 3 & 4 \\ \hline 
\multicolumn{5}{c}{} \\
\end{tabular}
\begin{tabular}{|c|c|c|c|} \hline
$a$ & $\lambda^{1}_{a} = \lambda^{2}_{a}$ &
$\mu^{1}_{a}$ &
$\mu^{2}_{a}$ \\ \hline \hline
1 & 0.166667 & 0 & 0 \\ \hline
2 & 0.333333 & -10.6667 & -1 \\ \hline
3 & 0.5 & -710.741 & -93.4444 \\ \hline
4 & 0.666667 & -26078.3 & -4678.83 \\ \hline
5 & 0.833333 & -424201 & -119094 \\ \hline
\multicolumn{4}{c}{$\gamma = 0.9$} \\
\end{tabular}
\caption{Arbitrary action selection}
\label{fig:2state-arbitrary}
\end{subfigure}

\begin{subfigure}{.3\textwidth}
\centering
\begin{tabular}{|c|c|c|c|c|}
\hline
0 & -- & -- & -- & -- \\ \hline    
1 & -- & -- & -- & -- \\ \hline    
2 & -- & -- & -- & -- \\ \hline    
3 & -- & -- & -- & -- \\ \hline    
4 & 5 & 6 & 7 & 8 \\ \hline 
\multicolumn{5}{c}{} \\
\end{tabular}

\begin{tabular}{|c|c|c|} \hline
$a$ & $\lambda^{1}_{a} = \lambda^{2}_{a}$ &
$\mu^{1}_{a} = \mu^{2}_{a} / 5$ \\ \hline \hline
1 & 1 & 0 \\ \hline
2 & 0.62548 & 0.761905 \\ \hline
3 & 0.872923 & 0.408163 \\ \hline
4 & 0.96675 & 0.172768 \\ \hline
5 & 1 & 0.0514189 \\ \hline
\multicolumn{3}{c}{$\gamma = 0.989716$} \\
\end{tabular}
\caption{Max-gain action selection}
\label{fig:2state-max-gain}
\end{subfigure}
}
\caption{(a) ``Transition probability, reward'' pairs for the transitions out of state 1 for action $a$, and for transitions out of state $2$ for action $a^{\prime}$. Note that there are $k$ such actions for each state (not all displayed). (b) Construction of a Hamiltonian path for $k = 5$.
Each cell in the matrix on top shows the index of the corresponding policy (interpreted as $\langle$row, column$\rangle$) in the visitation sequence $(0, 1, \dots, 24)$. The parameters listed below represent a single MDP from the feasible range to induce the Hamiltonian path above for $\mathcal{L}$. (c) For $k = 5$, the matrix on top shows a sequence of $9$ policies, beginning with $\langle 1, 1 \rangle$, which can be visited by a max-gain PI variant $\mathcal{L}$. Below are the parameters of an MDP on which this sequence is feasible. In this case, the final policy of the sequence, $\langle 5, 5 \rangle$ is an optimal policy for the MDP.
}
\label{fig:2state}
\end{figure}
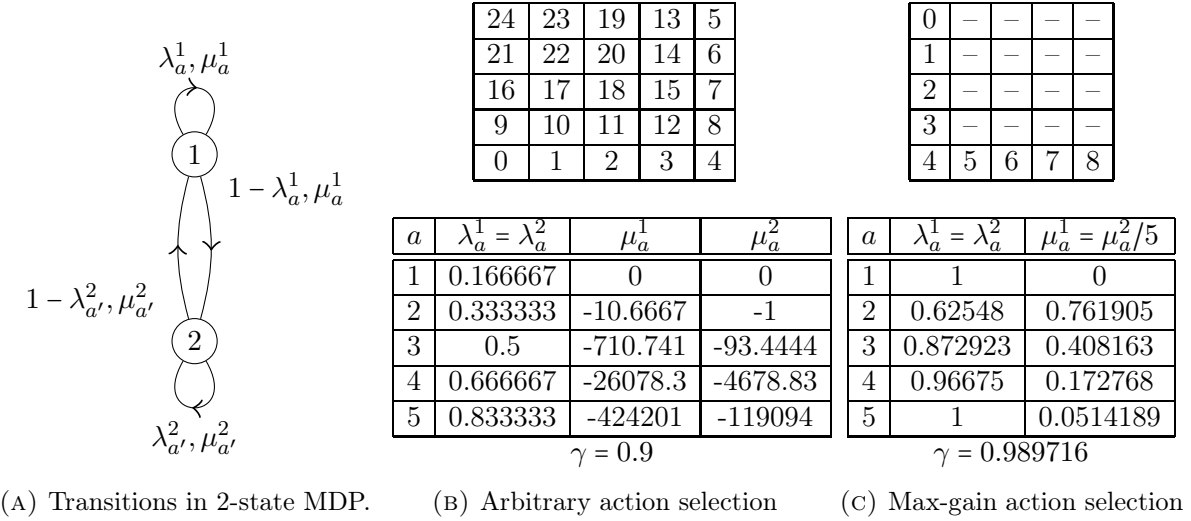

We describe families of $2$-state, $k$-action MDPs, $k \geq 2$, along with corresponding starting policies and PI algorithms, in order to establish the lower bounds in propositions 
\ref{prop:2state-arbitrary-mdp-lb} and
\ref{prop:2state-max-gain-mdp-lb}. In both constructions, we take the set of states to be $S = \{1, 2\}$, and the set of actions to be $A = \{1, 2, \dots, k\}$. We introduce simplifying notation for the transition probabilities and rewards. For $s \in S, a \in A$, let $\lambda^{s}_{a}$ denote the probability of transitioning from $s$ to $s$ by taking action $a$. Hence, if $T$ is the transition function, we have $T(s, a, s) = \lambda^{s}_{a}$ and $T(s, a, s^{\prime}) = 1 - \lambda^{s}_{a}$ for $s^{\prime} \in S \setminus \{s\}$. Similarly, for $s \in S, a \in A$, let $\mu^{s}_{a}$ denote the expected reward obtained by taking action $a$ from state $s$. Thus,  if $R$ is the reward function, we have $R(s, a) = \mu^{s}_{a}$. Figure~\ref{fig:lambda-mu} pictorially depicts state transitions in terms of  $\lambda^{\cdot}_{\cdot}$ and $\mu^{\cdot}_{\cdot}$. We continue to denote the discount factor $\gamma$. Since we only have two states, it is convenient to view policies as ordered pairs. For $i, j \in A$, we use $\langle i, j \rangle$ to denote the policy that takes action $i$ from state $1$ and action $j$ from state $2$. With this notation, the value function of policy $\langle i, j \rangle$ works out to
\begin{align}
V^{\langle i, j \rangle}(1) = \frac{\mu^1_i(1-\gamma \lambda^2_j) + \mu^2_j \gamma (1 - \lambda^1_i)}{(1 - \gamma)\{1 + \gamma(1 - \lambda^{1}_{i} - \lambda^{2}_{j})\}}; \text{    }
V^{\langle i, j \rangle}(2) = \frac{\mu^2_j(1-\gamma \lambda^1_i)+ \mu^1_i\gamma (1 - \lambda^2_j)}{(1 - \gamma)\{1 + \gamma(1 - \lambda^{1}_{i} - \lambda^{2}_{j})\}}. \label{eqn:2state-values}
\end{align}
Another useful function for our constructions is the gain function. For $i^{\prime}, j^{\prime} \in A$, we observe that
\begin{align}
\rho^{ \langle i, j \rangle}(1, i^{\prime}) = \mu^{1}_{i^{\prime}} - \mu^{1}_{i} + \frac{\gamma (\lambda^{1}_{i^{\prime}} - \lambda^{1}_{i}) (\mu^{1}_{i} - \mu^{2}_{j})}{1 + \gamma(1 - \lambda^{1}_{i} - \lambda^{2}_{j})}; \text{    }
\rho^{ \langle i, j \rangle}(2, j^{\prime}) = \mu^{2}_{j^{\prime}} - \mu^{2}_{j} - \frac{\gamma (\lambda^{2}_{j^{\prime}} - \lambda^{2}_{j}) (\mu^{1}_{i} - \mu^{2}_{j})}{1 + \gamma(1 - \lambda^{1}_{i} - \lambda^{2}_{j})}.\label{eqn:2state-gains}
\end{align}
Our constructions specify ranges for $\lambda^{s}_{a}$ and $\mu^{s}_{a}$ for $s \in S, a \in A$, and for the discount factor $\gamma$.

\subsubsection{Construction for Proposition~\ref{prop:2state-arbitrary-mdp-lb}.}
\label{subsubsec:proof-arbitrary-lb}

Whereas a $2$-state, $k$-action MDPs has $k^{2}$ policies, it only contains $\Theta(k)$ parameters in its transition and reward functions. To assemble our lower bound, we must simultaneously propose (1) a sequence in which the $k^{2}$ policies can be visited---non-trivial since some sequences are not feasible---and (2) a configuration of the MDP parameters so that this sequence is along a path in its PI-DAG. Below we describe one possible construction.\\

\noindent\textbf{Sequence of policies.} The sequence of policies we employ is best explained by arranging the $k^{2}$ policies in a matrix: for $i, j \in A$, the policy $ \langle i, j \rangle$ corresponds to the $i$-th row and $j$-th column of the matrix.
\begin{enumerate}
\item Our sequence begins with the policy $\pi_{0} = \langle k, 1 \rangle$.
\item Then we proceed along the last row, through policies $\langle k, 2\rangle, \langle k, 3 \rangle, \dots$, up to policy $ \langle k, k \rangle$.
\item Now keeping the column the same, we switch to the first row: that is, our next policy is $\langle 1, k \rangle$.
\item Thereafter, the sequence goes down the last column, through policies $\langle 2, k \rangle, \langle 3, k \rangle, \dots$, up to policy $\langle k - 1, k \rangle$.
\item From $\langle k - 1, k \rangle$, we again switch column and update to $\langle k - 1, 1 \rangle$. We are done if $k = 2$, else we go to step 1 and recursively cover the matrix formed by the first $k - 1$ rows and $k - 1$ columns. 
\end{enumerate}
Figure~\ref{fig:2state-arbitrary} shows the sequence of policies visited for $k = 5$. In general, we can obtain the next policy in the sequence for each policy $\langle i, j \rangle$, where $i, j \in A$ and $i \neq 1$ or $j \neq 1$, by calling the function

\begin{align*}
\text{next}(\langle i, j\rangle) &= \begin{cases}
\langle i, j + 1\rangle, & \text{if } i > j,\\
\langle 1, j \rangle, & \text{if } i = j,\\
\langle i + 1, j \rangle, & \text{if } i < j - 1,\\
\langle i, 1 \rangle, & \text{if } i = j - 1.\\
\end{cases}
\end{align*}

We take $\mathcal{L}$ to be a PI algorithm that chooses $\text{next}(\langle i, j \rangle )$ as the policy to follow $\langle i, j \rangle$ whenever possible; otherwise $\mathcal{L}$ may perform an arbitrary policy improvement.\\

\begin{center}
\framebox[0.76\textwidth]{
\parbox{0.67\textwidth}{
1. Select $\gamma$ arbitrarily from $(0, 1)$.\\
2. Select $\lambda^{1}_{1}$ and $\lambda^{2}_{1}$ arbitrarily from $(0, 1)$.\\
3. Set $\mu^{1}_{1}$ and $\mu^{2}_{1}$ to be arbitrary finite real numbers.\\
4. For $a = 1, 2, 3, \dots, k - 1$:\\
5. \phantom{aaaa}Select $\lambda^{1}_{a + 1}$ arbitrarily from $(\lambda^{1}_{a}, 1)$.\\
6. \phantom{aaaa}Select $\lambda^{2}_{a + 1}$ arbitrarily from $(\lambda^{2}_{a}, 1)$.\\
7. \phantom{aaaa}Set $\mu^{2}_{a + 1}$ to an arbitrary finite real number smaller than 
$$\min_{1 \leq i \leq a - 1} \left( 
\mu^{1}_{i} + \frac{(\mu^{1}_{i + 1} - \mu^{1}_{i})(1 + \gamma (1 - \lambda^{1}_{i} - \lambda^{1}_{a + 1}))}{\gamma (\lambda^{1}_{i + 1} - \lambda^{1}_{i})}
\right)$$
\phantom{7. aaaa} and also smaller than 
$$\frac{\mu^{2}_{1}(1 + \gamma (1 - \lambda^{1}_{1} - \lambda^{2}_{a + 1})) + \gamma (\lambda^{2}_{a + 1} - \lambda^{2}_{1}) \mu^{1}_{a}}{1 + \gamma (1 - \lambda^{1}_{1} - \lambda^{2}_{1})}.$$
8. \phantom{aaaa}Set $\mu^{1}_{a + 1}$ to an arbitrary finite real number smaller than $$\min_{1 \leq j \leq a} \left(\mu^{2}_{j} + \frac{(\mu^{2}_{j + 1} - \mu^{2}_{j})(1 + \gamma (1 - \lambda^{1}_{a + 1} - \lambda^{2}_{j}))}{\gamma (\lambda^{2}_{j + 1} - \lambda^{2}_{j})}\right)$$
\phantom{8. aaaa}and also smaller than
$$\frac{\mu^{1}_{1}(1 + \gamma (1 - \lambda^{1}_{a + 1} - \lambda^{2}_{a + 1})) + \gamma (\lambda^{1}_{a + 1} - \lambda^{1}_{1}) \mu^{2}_{a + 1}}{1 + \gamma (1 - \lambda^{1}_{1} - \lambda^{2}_{a + 1})}.$$
}
}
\end{center}

\noindent\textbf{MDP.} In the box above, we specify a procedure to set the parameters of an MDP such that $\text{next}( \langle i, j \rangle)$ necessarily dominates policy $ \langle i, j \rangle$ when $i \neq 1$ or $j \neq 1$. The policy $\langle 1, 1 \rangle$, on which the $\text{next}(\cdot)$ function has not been defined, is the unique optimal policy of our MDP. Notice in our construction that the sequences $(\lambda^{1}_{1}, \lambda^{1}_{2}, \dots, \lambda^{1}_{k})$ and $(\lambda^{2}_{1}, \lambda^{2}_{2}, \dots, \lambda^{2}_{k})$ are both monotonically increasing. The sequences can be constructed independently of each other, and also independent of the reward sequences $(\mu^{1}_{1}, \mu^{1}_{2}, \dots, \mu^{1}_{k})$ and $(\mu^{2}_{1}, \mu^{2}_{2}, \dots, \mu^{2}_{k})$. For $a \in A \setminus \{k\}$, (1) the acceptable range for $\mu^{2}_{a + 1}$ depends only on variables with indices at most $a$, apart from $\lambda^{1}_{a + 1}$ and $\lambda^{2}_{a + 1}$, and (2) the acceptable range for $\mu^{1}_{a + 1}$ depends only on variables with indices at most $a$, 
apart from $\lambda^{1}_{a + 1}$, $\lambda^{2}_{a + 1}$, and $\mu^{2}_{a + 1}$.


Figure~\ref{fig:2state-arbitrary} provides an example of an MDP constructed as above for $k = 5$. In this MDP, $\gamma$ is set to $0.9$, and for $a \in A$, $\lambda^{1}_{a} = \lambda^{2}_{a} = \frac{a}{k + 1}$. The rewards for action $1$---that is, $\mu^{1}_{1}$ and  $\mu^{2}_{1}$---are both set to $0$. For $a \in A \setminus \{k\}$, $\mu^{1}_{a + 1}$ and $\mu^{2}_{a + 1}$ are set to be $1$ less than than the smallest quantities on lines 7 and 8, respectively.

As we shall see next, the ranges for $\mu^{1}_{a + 1}$ and $\mu^{2}_{a + 1}$ are specified precisely so that the $\text{next}(\cdot)$ function always gives a legal policy improvement step.\\

\noindent\textbf{Proof.} Notice that $\text{next}( \langle i, j \rangle)$ always differs from $\langle i, j \rangle$ in exactly one action; it suffices for us to show that the different action 
in $\text{next}( \langle i, j \rangle)$ is an improving one for policy $\langle i, j \rangle$. Equivalently, we need to show that our construction satisfies the following constraints for $a \in A \setminus \{k\}$:
\begin{align}
\rho^{\langle a + 1, j \rangle}(2, j + 1) &> 0 &\text{for } 1 \leq j \leq a, \label{eqn:ham-c1}\\
\rho^{\langle a + 1, a +1 \rangle}(1, 1) &> 0, \label{eqn:ham-c2}\\
\rho^{\langle i, a + 1 \rangle}(1, i + 1) &> 0 &\text{for } 1 \leq i \leq a - 1. \label{eqn:ham-c3}\\
\rho^{\langle a, a + 1 \rangle}(2, 1) &> 0. \label{eqn:ham-c4}
\end{align}

We rewrite \eqref{eqn:ham-c1}--\eqref{eqn:ham-c4} based on the expansion of the gain function in \eqref{eqn:2state-gains}, and using the fact that 
$(\lambda^{1}_{1}, \lambda^{1}_{2}, \dots, \lambda^{1}_{k})$ and $(\lambda^{2}_{1}, \lambda^{2}_{2}, \dots, \lambda^{2}_{k})$ are monotonically increasing sequences. The four equivalent constraints, respectively, are, for $a \in A \setminus \{k\}$:
\begin{align}
\mu^{1}_{a + 1} &< \mu^{2}_{j} + \frac{(\mu^{2}_{j + 1} - \mu^{2}_{j})(1 + \gamma (1 - \lambda^{1}_{a + 1} - \lambda^{2}_{j}))}{\gamma (\lambda^{2}_{j + 1} - \lambda^{2}_{j})}, &\text{for } 1 \leq j \leq a,\label{eqn:ham-c5}\\
\mu^{1}_{a + 1} &< \frac{\mu^{1}_{1}(1 + \gamma (1 - \lambda^{1}_{a + 1} - \lambda^{2}_{a + 1})) + \gamma (\lambda^{1}_{a + 1} - \lambda^{1}_{1}) \mu^{2}_{a + 1}}{1 + \gamma (1 - \lambda^{1}_{1} - \lambda^{2}_{a + 1})},\label{eqn:ham-c6}\\
\mu^{2}_{a + 1} &< \mu^{1}_{i} + \frac{(\mu^{1}_{i + 1} - \mu^{1}_{i})(1 + \gamma (1 - \lambda^{1}_{i} - \lambda^{1}_{a + 1}))}{\gamma (\lambda^{1}_{i + 1} - \lambda^{1}_{i})}, &\text{for } 1\leq i \leq a - 1,\label{eqn:ham-c7}\\
\mu^{2}_{a + 1} &< \frac{\mu^{2}_{1}(1 + \gamma (1 - \lambda^{1}_{1} - \lambda^{2}_{a + 1})) + \gamma (\lambda^{2}_{a + 1} - \lambda^{2}_{1}) \mu^{1}_{a}}{1 + \gamma (1 - \lambda^{1}_{1} - \lambda^{2}_{1})}.\label{eqn:ham-c8}
\end{align}

The choice of $\mu^{2}_{a + 1}$ in our construction (line 8) ensures that \eqref{eqn:ham-c5} and \eqref{eqn:ham-c6} are satisfied, while that of $\mu^{1}_{a + 1}$ (on line 7) ensures that \eqref{eqn:ham-c7} and \eqref{eqn:ham-c8} are satisfied. It is due to our definition of the policy improvement sequence (the $\text{next}(\cdot)$ function) that the constraints on the MDP parameters are feasible; this is not necessarily true for other policy improvement sequences.

\subsubsection{Construction for Proposition~\ref{prop:2state-max-gain-mdp-lb}.}
\label{subsubsec:proof-max-gain-lb}

We follow the same steps as in Section~\ref{subsubsec:proof-arbitrary-lb} to show an $\Omega(k)$ lower bound for PI with max-gain action selection.\\

\noindent\textbf{Sequence of policies.} Our sequence of policies begins with $\langle 1, 1 \rangle$, and proceeds through policies $\langle 2, 1 \rangle, \langle 3, 1 \rangle, \dots$, up to $\langle k, 1 \rangle$. Thereafter we proceed through policies
$\langle k, 2 \rangle, \langle k, 3 \rangle, \dots$, up to policy $\langle k, k \rangle$. Hence, for this construction, we have
\begin{align*}
\text{next}(\langle i, j\rangle) &= \begin{cases}
\langle i + 1, j\rangle, & \text{if } i < k, j = 1,\\
\langle i, j + 1\rangle, & \text{if } i = k, j < k.\\
\end{cases}
\end{align*}

Notice that $\text{next}(\cdot)$ is only defined for a total of $2k - 2$ policies: those in which either the action at state $2$ is $1$, or the action at state $1$ is $k$ (but excluding $\langle k, k \rangle$, which is the last policy in our sequence). Figure~\ref{fig:2state-max-gain} shows the sequence of policies defined by $\text{next}(\cdot)$ for $k = 5$.

We take $\mathcal{L}$ to be a PI algorithm that chooses $\text{next}(\langle i, j \rangle )$ as the policy to follow $\langle i, j \rangle$ whenever possible; otherwise $\mathcal{L}$ may arbitrarily switch to the max-gain action at one or both states. We need to construct a $2$-state, $k$-action MDP in which the action to switch to according to $\text{next}( \langle i, j \rangle)$ (wherever applicable) is indeed a max-gain action.\\

\noindent\textbf{MDP.}  Let $\epsilon$ be an arbitrary
positive finite real number. In our construction, we set 
\begin{align*}
\gamma &= 1 - \frac{1}{k (2 + \epsilon)^{k - 1}},\\
\lambda^{1}_{a} &=
\begin{cases}
1, & \text{if } a = 1,\\
\frac{1}{\gamma} \left( 1 - \frac{k - a + 1}{k (2 + \epsilon)^{a - 1}} \right), & \text{if } a \in A \setminus \{1\},\\ 
\end{cases}\\
\lambda^{2}_{a} &= \lambda^{1}_{a} \text{ for } a\in A,\\
\mu^{1}_{a} &= 
\begin{cases}
0, & \text{if } a = 1,\\
\frac{a(k- a + 1)}{k (2 + \epsilon)^{a - 1}}, & \text{if } a \in A \setminus \{1\},
\end{cases}\\
\mu^{2}_{a} &= k \mu^{1}_{a} \text{ for } a\in A.
\end{align*}

For both states, action $1$ deterministically transitions to the same state, while meriting no reward. As we proceed through actions $2, 3, \dots, k$, the transition probabilities to the same state monotonically increase, while the rewards monotonically decrease (while remaining positive). These parameters are tuned so that the $1$-step reward among improving actions (which max-gain action selection greedily optimises)  is in the opposite order as the actions' long-term rewards. In Figure~\ref{fig:2state-max-gain}, we list the numerical values of the MDP parameters obtained for $k = 5$, by taking $\epsilon = 0.1$.\\

\noindent\textbf{Proof.} It suffices to show that our MDP implements the following constraints: (1) for $1 \leq a \leq k - 1$,
\begin{align}
\rho^{\langle a, 1 \rangle}(1, a + 1) &> 0, \label{eqn:mg-c1}\\
\rho^{\langle k, a \rangle}(2, a + 1) &> 0, \label{eqn:mg-c2}
\end{align}
and (2) for $1 \leq a \leq k - 2$, $a + 1 \leq i, j \leq k - 1$,
\begin{align}
\rho^{\langle a, 1 \rangle}(1, i) &> 
\rho^{\langle a, 1 \rangle}(1, i + 1), \label{eqn:mg-c3}\\
\rho^{\langle k, a \rangle}(2, j) &> 
\rho^{\langle k, a \rangle}(2, j + 1). \label{eqn:mg-c4}
\end{align}

Just as \eqref{eqn:mg-c1} and \eqref{eqn:mg-c3} show that the sequence of max-gain switches to state $1$ starting from $\langle 1, 1 \rangle$ takes us through $\langle 2, 1 \rangle, \langle 3, 1 \rangle, \dots, \langle k, 1 \rangle$, \eqref{eqn:mg-c2} and \eqref{eqn:mg-c4} show that the sequence of max-gain switches to state $2$ starting from $\langle k, 1 \rangle$ takes us through $\langle k, 2 \rangle, \langle k, 3 \rangle, \dots, \langle k, k \rangle$. \eqref{eqn:mg-c2} shows that $a + 1$ is an improving action at state $2$ for policy $\langle k, a \rangle$, $1 \leq a \leq k - 1$, while \eqref{eqn:mg-c4} establishes that it is indeed the max-gain action at state $2$ for policy $\langle k, a \rangle$ (from the improving set $\{a + 1, a + 2, \dots, k\}$). We prove \eqref{eqn:mg-c1}, \eqref{eqn:mg-c2}, \eqref{eqn:mg-c3}, and \eqref{eqn:mg-c4} to complete our proof. In each case we expand the corresponding gain terms as provided in \eqref{eqn:2state-gains}.

\eqref{eqn:mg-c1} shows that $a + 1$ is an improving action at state $1$ for policy $\langle a, 1 \rangle$, $1 \leq a \leq k - 1$. By implication: $\langle 1, 1 \rangle \prec \langle 2, 1 \rangle \prec \dots \prec \langle k, 1 \rangle$. Applying Theorem~\ref{thm:policy-improvement} to policies in this sequence, we observe that for policy $\langle a, 1 \rangle$, no action from the set $\{1, 2, \dots, a - 1\}$ can be an improving action at state $1$. \eqref{eqn:mg-c3} establishes that among the remaining actions, $a + 1$ is indeed the one with the maximum gain at state $1$.
Just as \eqref{eqn:mg-c1} and \eqref{eqn:mg-c3} show that the sequence of max-gain switches to state $1$ starting from $\langle 1, 1 \rangle$ takes us through $\langle 2, 1 \rangle, \langle 3, 1 \rangle, \dots, \langle k, 1 \rangle$, \eqref{eqn:mg-c2} and \eqref{eqn:mg-c4} show that the sequence of max-gain switches to state $2$ starting from $\langle k, 1 \rangle$ takes us through $\langle k, 2 \rangle, \langle k, 3 \rangle, \dots, \langle k, k \rangle$. \eqref{eqn:mg-c2} shows that $a + 1$ is an improving action at state $2$ for policy $\langle k, a \rangle$, $1 \leq a \leq k - 1$. By Theorem~\ref{thm:policy-improvement}, no action from the set $\{1, 2, \dots, a - 1\}$ can be an improving action at state $2$ for policy $\langle k, a\rangle$. \eqref{eqn:mg-c4} establishes that among actions from the set $\{a + 1, a + 2, \dots, k\}$, indeed $a + 1$ is the one with the maximum gain at state $2$ for policy $\langle k, a \rangle$. We prove \eqref{eqn:mg-c1}, \eqref{eqn:mg-c2}, \eqref{eqn:mg-c3}, and \eqref{eqn:mg-c4} to complete our proof. In each case we expand the corresponding gain terms as provided in \eqref{eqn:2state-gains}.

As proof of \eqref{eqn:mg-c1}, observe that for $1 \leq a \leq k - 1$, $$\rho^{\langle a, 1 \rangle}(1, a + 1) = \frac{k - a}{k (2 + \epsilon)^{a}} > 0.$$ As proof of \eqref{eqn:mg-c2}, we obtain  that for $1 \leq a \leq k - 1$, $$\rho^{\langle k, a \rangle}(2, a + 1) = \begin{cases}
\mu^{2}_{2} + \frac{\gamma}{1 - \gamma} (1 - \lambda^{2}_{2})\mu^{1}_{k}, & \text{if } a = 1,\\ 
\frac{(2 + \epsilon)^{k - a - 1} a (k - a) - 1}{(2 + \epsilon)^{k - 1}} + \frac{k - 1}{(k - a + 1)(2 + \epsilon)^{k}}, & \text{if } 2 \leq a \leq k - 1,\\
\end{cases}$$ which in both cases is seen to be a sum of positive terms. As proof of \eqref{eqn:mg-c3}, notice that for $1 \leq a \leq k - 2, a + 1 \leq i \leq k - 1$, $$\rho^{\langle a, 1 \rangle}(1, i) - \rho^{\langle a, 1 \rangle}(1, i + 1) = \frac{\{(1 + \epsilon)(i - a) - 1\}(k - i)}{{k (2 + \epsilon)^{i}}} + \frac{(2 + \epsilon)(i - a)}{k (2 + \epsilon)^{i}} > 0.$$ As proof of \eqref{eqn:mg-c4}, we have for $1 \leq a \leq k - 2, a + 1 \leq j \leq k - 1$, $$\rho^{\langle k, a \rangle}(1, j) - \rho^{\langle k, a \rangle}(1, j + 1) = \frac{\{(1 + \epsilon)(kj - a) - k\}(k - j)}{{k (2 + \epsilon)^{j}}} + \frac{(2 + \epsilon)(j - a)}{k (2 + \epsilon)^{i}} + \frac{\gamma(\lambda^{2}_{j + 1} - \lambda^{2}_{j})\mu^{1}_{k}}{1 - \gamma \lambda^{2}_{a}} > 0.$$

\subsection{Proofs of upper bounds for DMDPs.}
\label{subsec:2statebounds-proofsofupperboundsfordmdps}

We again take $S = \{1, 2\}$ and $A = \{1, 2, \dots, k\}$ for an arbitrary DMDP $M$. Since transitions are deterministic in $M$, for each state we can partition $A$ into two subsets: actions that transition into the same state, and those that transition into the other state. Concretely, let $A^{1}_{\text{same}} \subseteq A$ be the set of actions which when applied at state $1$, transition into state $1$. In other words, $A^{1}_{\text{same}}(1) = \{a \in A, \lambda^{1}_{a} = 1\}$. Also define $A^{1}_{\text{other}} = A \setminus A^{1}_{\text{same}}$.

Let $\pi_{0}, \pi_{1}, \dots, \pi_{\ell}$ be the sequence of policies visited by PI algorithm $\mathcal{L}$ on $M$, if initialised at an arbitrary policy $\pi_{0}$. In this sequence of policies, we focus on the sequence of switches to the action at state $1$. Let $a_{0}$ be the action in the initial policy $\pi_{0}$; let $a_{1}$ be the action to which $a_{0}$ is switched; let $a_{2}$ be the action to which $a_{1}$ is switched; and so on. Let there be $\ell^{1}$ switches to state $1$ in total. Note that the sequence of actions $(a_{0}, a_{1}, a_{2}, \dots, a_{\ell^{1}})$ need \textit{not} be in 1-to-1 correspondence with $(\pi_{0}, \pi_{1}, \pi_{2}, \dots, \pi_{\ell}$), since some iterations may only switch the action at state $2$. 

\subsubsection{Proof of Proposition~\ref{prop:2state-arbitrary-dmdp-ub}.}

We observe two constraints on $(a_{0}, a_{1}, a_{2}, \dots, a_{\ell^{1}})$.

First, any action $a \in A^{1}_{\text{same}}$ can occur at most once in $(a_{0}, a_{1}, a_{2}, \dots, a_{\ell^{1}})$. The reason is as follows. The value of state $1$ under any policy $\pi$ such that $\pi(1) = a$ is exactly $\frac{R(1, a)}{1 - \gamma}$. In particular, since $a \in A^{1}_{\text{same}}$, $V^{\pi}(1)$ does not depend on $\pi(2)$. Now, if $a \in A^{1}_{\text{same}}$ is switched to an improving action, it means that state $1$ should get a strictly higher value in the resulting policy (as evident from Theorem~\ref{thm:policy-improvement}). Since values cannot decrease under PI, state $1$ cannot once again get $a$ as its action, since that would return its value to $\frac{R(1, a)}{1 - \gamma}$.

Second, in between any two occurrences of an action $a \in A^{1}_{\text{other}}$ in $(a_{0}, a_{1}, a_{2}, \dots, a_{\ell^{1}})$, there must be some action $a^{\prime} \in A^{1}_{\text{same}}$. To see why, consider an arbitrary contiguous subsequence of actions $a_{i}, a_{i + 1}, \dots, a_{j}$ for some $0 \leq i < j \leq \ell^{1}$, in which every action is from $A^{1}_{\text{other}}$. Since $a_{i + 1}$ is an improving action for state $1$ for some policy $\pi$ satisfying $\pi(1) = a_{i}$, we must have $R(1, a_{i + 1}) > R(1, a_{i})$. By the same argument, $R(1, a_{i + 2}) > R(1, a_{i + 1})$, $R(1, a_{i + 3}) > R(1, a_{i + 2})$, and so on. Due to this monotonic increase in rewards within the contiguous subsequence, we cannot have $a_{j} = a_{i}$.

In summary, we have that any action $a \in A^{1}_{\text{same}}$ can occur at most once in $(a_{0}, a_{1}, a_{2}, \dots, a_{\ell^{1}})$, and any action $a \in A^{1}_{\text{other}}$ can occur at most $|A^{1}_{\text{same}}| + 1$ time(s) in $(a_{0}, a_{1}, a_{2}, \dots, a_{\ell^{1}})$. Consequently, the length of the sequence $(a_{0}, a_{1}, a_{2}, \dots, a_{\ell^{1}})$ is

\begin{align*}
\ell^{1} + 1 &\leq \left|A^{1}_{\text{other}}\right| \left(\left|A^{1}_{\text{same}}\right| + 1\right) +  \left|A^{1}_{\text{same}}\right|
= 
\left|A^{1}_{\text{other}}\right| (k - \left|A^{1}_{\text{other}}\right| + 1) +  k - \left|A^{1}_{\text{other}}\right|\\
&= \frac{k^{2}}{4} + k - \left(\frac{k}{2} - \left|A^{1}_{\text{other}}\right|\right)^{2} \leq \frac{k^{2}}{4} + k.
\end{align*}

The number of switches done to state $1$ is $\ell^{1} \leq \frac{k^{2}}{4} + k - 1$. We may repeat our argument to claim the same upper bound on the number of switches done to state $2$. Since at least one action is switched in each improvement made by PI, the total number of policies visited is at most $2(\frac{k^{2}}{4} + k - 1) + 1$, as claimed in the proposition.

\subsubsection{Proof of Proposition~\ref{prop:2state-max-gain-dmdp-ub}.}

To analyse max-gain action selection, we consider the sequence of ``switched-to'' actions $(a_{1}, a_{2}, \dots, a_{l^{1}})$. We note two constraints on this sequence.

First, there can be at most one action from $A^{1}_{\text{same}}$ that occurs in $(a_{1}, a_{2}, \dots, a_{l^{1}})$. Let us refer to $a \in A^{1}_{\text{same}}$ as a ``maximal same-action'' if $R(1, a) = \max_{a^{\prime} \in A^{1}_{\text{same}}} R(1, a^{\prime})$. Since the max-gain rule is used for action selection, it is immediately seen that any action $a \in A^{1}_{\text{same}}$ that occurs in $(a_{1}, a_{2}, \dots, a_{l^{1}})$ must be a maximal same-action. If $\pi(1) = a$ for a maximal same-action $a$, we have $V^{\pi}(1) = \max_{a^{\prime} \in A^{1}_{\text{same}}} \frac{R(1, a^{\prime})}{1 - \gamma}$. Since switching out a maximal same-action must increase the value at state $1$, no subsequent switch can be made to an action from $A^{1}_{\text{same}}$, as that would necessarily then decrease the value of state $1$ to $\max_{a^{\prime} \in A^{1}_{\text{same}}} \frac{R(1, a^{\prime})}{1 - \gamma}$.

Second, no two consecutive actions in $(a_{1}, a_{2}, \dots, a_{l^{1}})$ can both belong to $A^{1}_{\text{other}}$. Let us refer to $a \in A^{1}_{\text{other}}$ as a ``maximal other-action'' if $R(1, a) = \max_{a^{\prime} \in A^{1}_{\text{other}}} R(1, a^{\prime})$. Since the max-gain rule is used for action selection, any action $a \in A^{1}_{\text{other}}$ that occurs in $(a_{1}, a_{2}, \dots, a_{l^{1}})$ must be a maximal other-action. Clearly no action $a^{\prime} \in A^{1}_{\text{other}}$ can be an improving action for state $1$ in policy $\pi$ if $\pi(1)$ is a maximal other-action.

From the two constraints described above, it is clear that under max-gain action selection, $(a_{1}, a_{2}, \dots, a_{l^{1}})$ can have at most one action from $A^{1}_{\text{same}}$, and at most $2$ actions from $A^{1}_{\text{other}}$. For the same reasons, state $2$ can also be switched at most three times. Hence $\mathcal{L}$ can evaluate at most $2 \times 3 + 1 = 7$ policies.

%% file: conclusion.tex
\section{Summary and outlook}

We have presented a new perspective on the policy space of DMDPs, which yields running-time upper bounds that apply to the entire family of PI algorithms. We obtain an even tighter upper bound for the family of PI algorithms that switch only to max-gain actions. Central to our analysis is the set of cycles in directed multigraphs induced by DMDPs (Madani~\cite{Madani}, Post and Ye~\cite{PostYe}). Our core results are upper bounds on the number of cycles in certain families of induced multigraphs, which are defined based on the properties of corresponding PI algorithms. Considering $2$-state MDPs as a special case, we also show that PI can complete in strictly fewer iterations when transitions are constrained to be deterministic.

Our results give rise to several questions for further investigation.

\begin{itemize}
\item For the case of $k = 2$ actions, Melekopoglou and Condon~\cite{MelekopoglouCondon} furnish an MDP and a PI variant that can visit \textit{all} the $2^{n}$ policies for the MDP. It remains unknown if such a Hamiltonian path through the set of policies exists for any MDP with $k \geq 3$ actions. Interestingly, the tightest lower bound known for MDPs with $k \geq 3$ actions---of $\Omega(k^{n/2})$ iterations (Ashutosh~\emph{et al.}~\cite{Ashutosh})---arises from a DMDP. There appears to be room to obtain a tighter lower bound for MDPs with $k \geq 3$ actions. A concrete first step in pursuit of a ``Hamiltonian'' lower bound could be to attempt constructing a $3$-state, $3$-action MDP in which all $27$ policies can be visited by PI.

\item Our upper bounds for PI on DMDPs depend directly on our upper bounds on the number of path-cycles in induced multigraphs. Our current analysis only utilises the fact that a path-cycle, once replaced, cannot appear again. Tighter upper bounds might be provable by exploiting additional constraints. For example, it might be possible to argue that path-cycles $P_{1}$ and $P_{2}$ mutually exclude each other in the trajectory taken by PI since they necessarily induce incomparable value functions. Similarly, if it can be shown that 
there is no legal sequence of switches to go from any policy containing path-cycle $P_{1}$ to any policy containing path-cycle $P_{2}$, it would imply that only one of $P_{1}$ and $P_{2}$ can be visited by PI.
 
\item In Section~\ref{sec:2statebounds},  we have provided a $2$-state, $k$-action MDP, $k \geq 2$, on which a max-gain PI variant visits $\Omega(k)$ policies. This construction can easily be extended to $n$-state, $k$-action MDPs, $n \geq 2$, $k \geq 2$, such that a max-gain PI variant would visit $\Omega(nk)$ policies. The generalisation would merely require making $n - 1$ replicas of state $1$ in our current construction. It would be interesting to examine whether the dependence on $k$ in this lower bound can be made super-linear.

\item A standard approach to solve a given MDP is to solve an induced LP. For an $n$-state, $k$-action MDP, the standard ``primal'' LP formulation is based on $n$ variables and $nk$ constraints, while the ``dual'' is based on $nk$ variables and $n$ constraints (Puterman~\cite[see Section 6.9]{Puterman}). Some authors (Post and Ye \cite[see Section 2]{PostYe}) refer to the latter formulation as the primal and to the former as the dual---in what follows, we stick to the convention of Puterman~\cite{Puterman}.

Vertices in the dual LP polytope are in 1-to-1 correspondence with the policies of the MDP. Simplex algorithms to solve the dual LP may be viewed as PI variants that only perform a switch at a single state in each iteration (multi-switch PI amounts to block-pivoting Simplex updates).

The ``LP digraph'' of a linear program has vertices corresponding to the vertices of the feasible polytope, and directed edges from vertices to neighbours that improve the objective function. The LP digraph of the dual LP induced by an MDP is a subgraph of the PI-DAG of the underlying MDP. Hence, our upper bounds on path lengths in the PI-DAG of DMDPs also apply to paths in the induced dual LP digraph. Notably, the \textit{primal} LP induced by a DMDP contains at most two variables per constraint. LPs with this property have been well-studied; for instance, it is known that they can be solved in strongly polynomial time (Megiddo~\cite{Megiddo:1983}). Hence, our results for DMDPs can possibly be generalised to a larger class of LPs. In particular, one could attempt to furnish non-trivial upper bounds on path lengths in the dual LP digraph when the primal is restricted to having at most two variables per constraint.

\end{itemize}